\documentclass[useAMS,usenatbib,babel]{mn2e}
\usepackage[english,english]{babel}
\usepackage{amsmath}
\usepackage{amssymb,amsfonts,textcomp}
\usepackage{hyperref}
\usepackage[usenames]{color}
\hypersetup{dvips, colorlinks=true, linkcolor=black, citecolor=black, filecolor=black, urlcolor=black}
\usepackage{graphicx}
\usepackage{caption}
\usepackage{setspace}
\usepackage{multirow,balance}
\usepackage{todonotes}
\usepackage{soul}

\definecolor{grey}{rgb}{0.75,0.75,0.75}
\definecolor{Orange}{rgb}{1.0,0.5,0.15}
\definecolor{brown}{rgb}{0.7,0.25,0.0}
\definecolor{pink}{rgb}{1.0,0.5,0.5}
\definecolor{darkerred}{rgb}{0.8,0,0}
\definecolor{darkerblue}{rgb}{0,0,0.8}
\definecolor{Blue}{rgb}{0,0.08,0.65}
\definecolor{Red}{rgb}{0.65,0.08,0.05}
\definecolor{Green}{rgb}{0.15,0.45,0.25}

\def\gtrsim{\lower.5ex\hbox{$\; \buildrel > \over \sim \;$}}
\usepackage{graphicx}

\newcommand{\mn}{\mbox{{\sc \small Horizon}-MareNostrum\,\,}}

\begin{document}

\author[C. Laigle et al. ]{C. Laigle$^{1}$, C. Pichon$^{1,2}$, S. Codis$^{1}$, Y. Dubois$^{1}$,  D. Le Borgne$^{1}$, D. Pogosyan$^{3}$,\newauthor
J. Devriendt$^{4}$, S. Peirani$^{1}$, S. Prunet$^{1}$,  S. Rouberol$^{1}$,  A. Slyz$^{5}$,  T. Sousbie$^{1}$ \\
$^{1}$ Institut d'Astrophysique de Paris \& UPMC (UMR 7095), 98 bis boulevard Arago, 75014 Paris, France\\
$^{2}$ Institute of Astronomy, University of Cambridge, Madingley Road, Cambridge, CB3 0HA, UK\\
$^{3}$ Department of Physics, University of Alberta, 11322-89 Avenue, Edmonton, Alberta, T6G 2G7, Canada.\\
$^{4}$ Astrophysics, University of Oxford, Keble Road, Oxford OX1 3RH, UK\\
$^{5}$ KITP Kohn Hall-4030 University of California Santa Barbara, CA 93106-4030
 }
\title[Swirling filaments]{
{Swirling around filaments: are large-scale structure vortices spinning up  dark haloes?}
}
\date{Accepted 2014 October 29.  Received 2014 October 22; in original form 2013 October 13}

\maketitle

%-----------------------
\begin{abstract}
The kinematic analysis of dark matter and hydrodynamical simulations suggests that  the vorticity in large-scale structure is mostly confined to,  and predominantly aligned with their filaments, with an excess of probability of 20 per cent to have the angle between vorticity and filaments direction lower than 60$^{\rm o}$ relative to random orientations.
The cross-sections of these filaments are typically partitioned  into {four quadrants with opposite vorticity sign}, arising from multiple flows, originating from  neighbouring walls.
The spins of haloes embedded within these filaments are consistently aligned with this vorticity for any halo mass, with a stronger alignment for the most massive structures up to an excess of probability of 165 per cent. \\
  The global geometry of the flow within the cosmic web is therefore qualitatively consistent with a spin acquisition for smaller haloes induced by this large-scale  coherence, as argued in Codis et al.  
 In effect,  secondary anisotropic infall (originating from the vortex-rich  filament within which these lower-mass haloes form) dominates  the AM budget of these haloes.
 The transition mass from alignment to orthogonality is  related to the size of a given multi-flow region with a given polarity.
 This transition may  be reconciled with the standard tidal torque theory if the latter  is augmented so as to account  for the larger scale anisotropic environment of walls and filaments.
\end{abstract}
%-----------------------

\begin{keywords}
method: numerical -- galaxies: formation -- galaxies: haloes -- large-scale structure of Universe
\end{keywords}

%%%%%%%%%%%%%%%
\section{Introduction}
%%%%%%%%%%%%%%%
%

The standard paradigm of galaxy formation addresses the acquisition of spin via the so-called Tidal Torque Theory \citep[TTT][]{Hoyle49,peebles69,TTTR,TTT} for which collapsing protogalaxies acquire their spin because of a misalignment between their inertia tensor and their (local) tidal tensor. 
There is ample evidence that for massive (quasi-linear) clusters, TTT provides a sound theoretical framework in which to describe angular momentum (AM) acquisition during the linear phase of structure formation. 
Conversely, lighter non-linear structures undergo significant drift within the large-scale tidal field, and move some distance away from their original Lagrangian  patch \citep[see e.g.][for a review]{schaefer08}.  
Over the last 10 years, numerical simulations as well as theoretical consideration \citep{katz03,birnboim03,keresetal05,ocvirk08} have accumulated evidence 
that    the intricate cosmic web plays a critical role in the process of forming high-redshift galaxies.
In the initial phase of galaxy formation, the condition of the intergalactic medium  leads to essentially isothermal shocks. Hence cold gas follows closely the cosmic web while radiating away the thermal energy gained by the extraction of  kinetic energy every time its trajectory dictates the formation of a shock. 

The dynamical relevance of the anisotropy of the cosmic web for galaxy formation may have been  partially underestimated given the small mass involved (in contrast to the mass in peaks). Indeed, spherical collapse and Press--Schechter theory have been quite successful at explaining the mass function of galaxies \citep{PS1974}.
On the other hand, the morphology of galaxies, arguably a secondary feature, is controlled at high redshift by their spin (see e.g. \citealp{Dubois2011}) and is very likely driven by later infall of AM-rich gas.
 In turn, the critical ingredient must therefore be the anisotropy of such infall, driven by its dynamics within the cosmic web, which differ significantly (via the hitherto mentioned shocks) from that of the dark matter (DM), since cold flows advect the AM they acquired as they formed during the early phase of large-scale structure (LSS) formation. 
{A  paradigm for the acquisition of disc AM via filamentary flows was  recently  proposed by \citet{pichon11} which found a closer connection between the 3D geometry and dynamics of the neighbouring cosmic web and the properties of embedded dark haloes and galaxies than originally suggested by the standard hierarchical formation paradigm
 \cite[see also][]{Prieto2013,Stewart2013}.
 At these scales, in the surrounding asymmetric gravitational patch gas  streams out from the neighbouring voids, towards their encompassing filaments where it shocks, until the cold flows are swallowed by the forming galaxy, advecting their newly acquired AM  \citep{kimm11,tillson12}.  While the gas is streamed  out of the walls towards their
 surrounding filaments
 it winds up  and forms the first generation of galaxies with a spin parallel to the filaments 
 (\citealp{Calvo07,hahn07,Paz08,Zhang2009,codis12,libeskind13a}, \citealp[see also][for earlier indication of anisotropic inflows]{aubert04,b24}). 
These authors  explored the link between DM haloes' spins and the cosmic web to quantify this alignment.} They  detected a redshift-dependent mass transition $M_{\rm crit}$, varying with the scale (or equivalently with the hierarchical level of the cosmic structure in which the halo is embedded; see \citealp{Calvo13}). \cite{codis12} interpreted the correlation in terms of large-scale cosmic flows: high-mass haloes have their spins perpendicular to the filament because they are the results of major mergers \citep[see also][]{Peirani2004}; low-mass haloes are not the products of merger and acquire their mass by accretion, which explains that their spins are parallel to the filament. 
  \cite{danovichetal11} also studied the feeding of massive galaxies at high redshift through cosmic streams using the \mn simulation \citep{ocvirk08} and
   found that  galaxies are fed by one dominant stream with a tendency to be fed by three major streams.
All these investigations suggest the existence of an additional mechanism  affecting first low-mass haloes:   mass accretion in an anisotropic, multi-flow  environment. 

\cite{Tempel13} and \cite{Zhang2013} have recently found evidence of such alignment in the Sloan Digital Sky Survey (an orthogonality for S0 galaxies and a weak alignment for late-type spirals). 
\begin{table*}
\singlespacing
\begin{tabular}{cccccccc}
 \textbf{Name} & \textbf{Type} & \textbf{Box size}& \textbf{Resolution}& \textbf{R$_{\rm velocity}$}& \textbf{R$_{\rm density}$}& \textbf{R$_{\rm Lagrangian}$}&\textbf{Minimum halo mass}\\
  & & $h^{-1}\, \rm Mpc$ &&$h^{-1}\, \rm Mpc$&$h^{-1}\, \rm Mpc$&$h^{-1}\, \rm Mpc$& $10^{10}M_{\odot}$\\\hline
  $ \displaystyle {\cal S}^{ \rm CDM}_{100}$ & $\Lambda$CDM& 100 & $256^3$&0.39&2.3&--&44 \\
${\cal S}^{\rm HDM}_{100}$ & $\Lambda$HDM &100 & $256^3$&0.39&2.3&2.3&-- \\
 ${\cal S}^{\rm CDM}_{50}$ & $\Lambda$CDM & 50 & $256^3 \times (20)$ &0.78&1.2&--&6.2\\
${\cal S}^{\rm CDM}_{20}$ & $\Lambda$CDM &20 & $512^3$&0.039&0.23&-- &0.044\\
${\cal S}^{\rm HDM}_{20}$ & $\Lambda$HDM &20 & $512^3$ &0.039&0.23&0.23&--\\
${\cal S}^{\rm CDM}_{2000}$ & $\Lambda$CDM& 2000 & $4096^3$&--&5&--&77 \\
\end{tabular}
\caption{The set of simulations used in Sections~\ref{sec:result} and \ref{sec:interpretation}.
The so-called $\Lambda$HDM subset corresponds to simulations, the initial condition of which have been smoothed over 2.3 $h^{-1}\, \rm Mpc$ and 0.23 $h^{-1}\, \rm Mpc$.  The simulation ${\cal S}^{\rm CDM}_{2000}$ corresponds to the post-processing of  an HPC simulation which allowed us to identify over 34 million haloes. The velocity field, density field, initial conditions were smoothed with Gaussian filter. In this work, we consider haloes with more than 100 particles.
}
\label{tab:simu}
\end{table*}
%%%
The detailed origin of this correlation, while not strictly speaking surprising, as well as its measured dependence on mass, has not yet been fully understood. The spin of the dark halo represents, in essence, the vortical motion of the matter and as such can be expected to reflect the vorticity  in the surrounding protogalactic patch of a forming halo. 
%%%%
Indeed, to understand this trend, \cite{libeskind13b} have argued that the local {\sl vorticity} was more relevant  than 
 the original tidal field in setting up the direction of dark halo spins. 
They have  explored the link between vorticity in halo environment and the origin of haloes spin and found a strong alignment between both. 
Vorticity tends to be perpendicular to the axis along which material is collapsing fastest. 
A natural tell-tale of such process would be a significant large-scale vorticity generation {\sl in} the multiflow regions corresponding to the interior of filaments.
Recently, \citet{Wang13} revisited this description by introducing three invariants of the velocity gradient tensor and concluded that vorticity generation is highly correlated with large-scale structure before and after shell-crossing, in a way which depends on the flow morphology.
%%%
Vorticity arises only after shell crossing in multi-streaming regions and requires the look inside such regions.  Pioneering study of \cite{pichon99} theoretically demonstrated that in the simplest pancake-like  multistream collapse the level of the vorticity generated is of the order of Hubble constant at the collapse stage at the scale of the thickness of the forming structures.
%%%%
While relying on these theoretical predictions, \cite{codis12} speculated that secondary shell-crossing could lead to the formation of vortices aligned with the forming filament. In turn, these vortices would account for the spin  of these haloes.
 There is  now indeed ample numerical evidence that the evolution of galaxy morphology is  likely to be in part driven by the geometry of the cosmic web,
and in particular its vorticity content.

Hence our focus will be in revisiting these findings with an emphasis on 
{\sl where} (tracing the filaments) and {\sl why} (studying the origin of the vorticity and its orientation) these trends are detected. 
We will also tentatively explain the origin of the mass transition for halo--spin alignment with the LSS's filaments.
This paper aims at revisiting  early stages of AM acquisition corresponding to when the cold gas/DM is expelled from neighbouring voids and walls. The main question addressed in this work will be: are there statistical evidence that swirling filaments are responsible for spinning up dark haloes and gaseous discs?

The  focus will be specifically exclusively on lower mass haloes ($M_{\odot}<5 \,10^{12}$) for which secondary anisotropic infall (originating from the vortex-rich  filament within which they form) dominates  their angular  momentum budget.
To that end, we will in particular make use of filament and wall tracers in order to quantify the cosmic web,  which is the natural metric for galactic evolution.
The virtual data used will be DM and hydrodynamical simulations. 

This paper is organized as follows.
Section~\ref{sec:simu} describes the simulations and the estimators implemented  in this paper. 
Section~\ref{sec:result} sums up robust statistical results of  (i) the  orientation  of the vorticity relative to 
the filaments, (ii) the distribution of the vorticity inside the filament and (iii) the alignment of the spin of dark haloes with the vorticity.
Section~\ref{sec:interpretation}  explores qualitatively the origin of this vorticity 
and uses the link between vorticity and spin to explain the non-monotonic behaviour of spin--filament alignment for haloes with masses lower than $M_{\rm crit}$.
Section~\ref{sec:conclusion} wraps up and discusses implications.
Appendix~\ref{sec:gas-LSS}  finds consistency in alignment with 
the vorticity of adiabatic/cooling gas.
Appendix~\ref{sec:Toy} illustrates the transition mass in the spin/filament alignment via a simple toy model for the typical vorticity within the caustic.
Appendix~\ref{sec:persistence} studies the effect of persistence and Appendix~\ref{sec:smoothing} the effect of the variation of the smoothing scale on the alignment.
Appendix~\ref{sec:validation} analyses the orientation of the vorticity with respect to the tidal field eigendirections.
Appendix~\ref{sec:Fof} explains the cleaning of the Friend-of-Friend halo catalogue.

%%%%%%%%%%%
\section{Datasets and estimators}\label{sec:simu}
%%%%%%%%%%%%

\begin{figure*}
\includegraphics[scale=0.5]{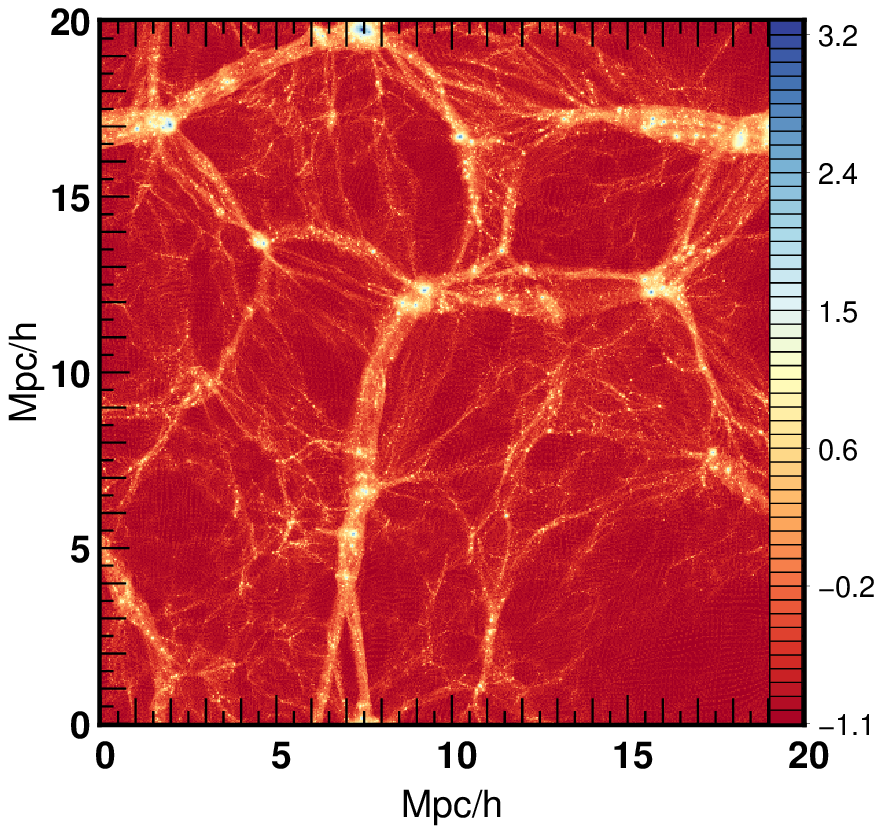}
\includegraphics[scale=0.5]{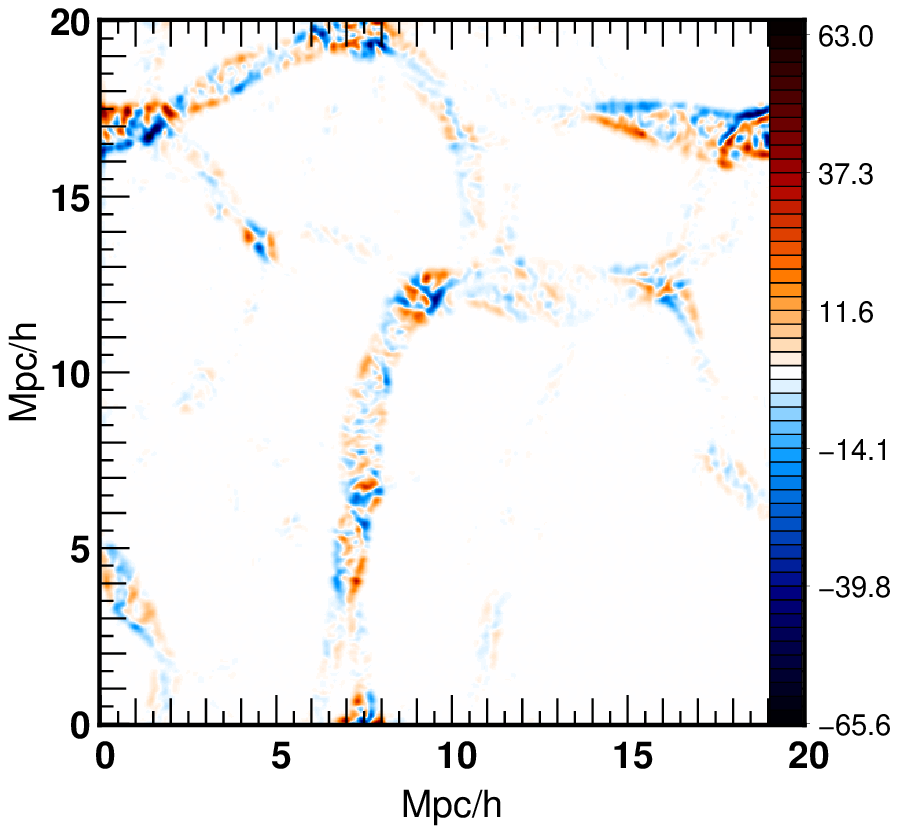}
  \includegraphics[scale=0.5]{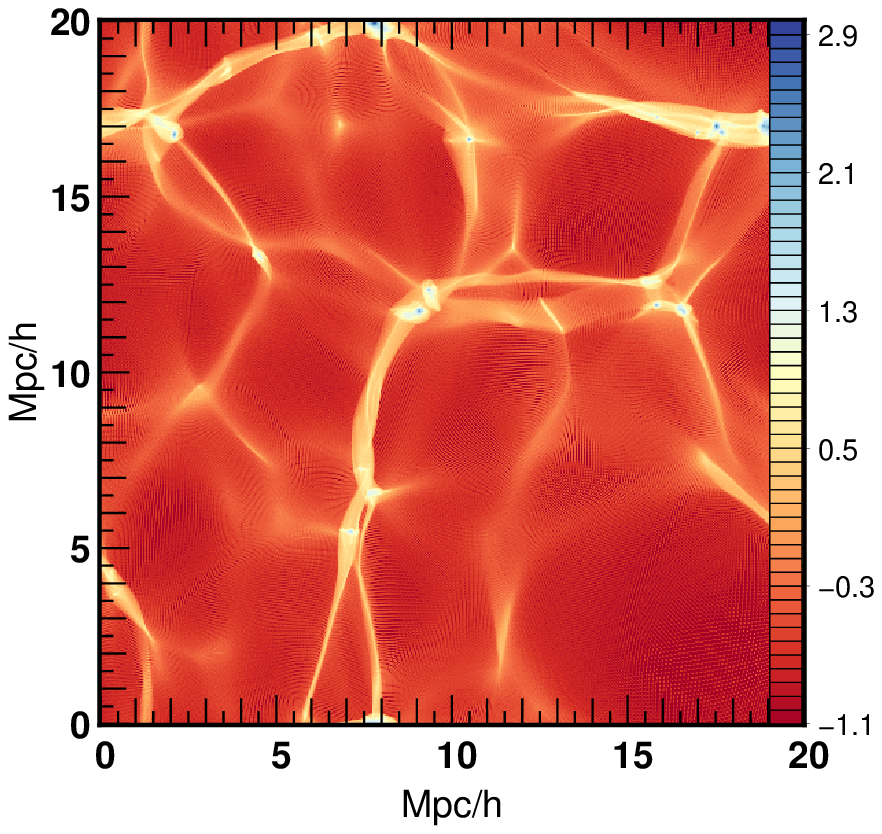}
\includegraphics[scale=0.5]{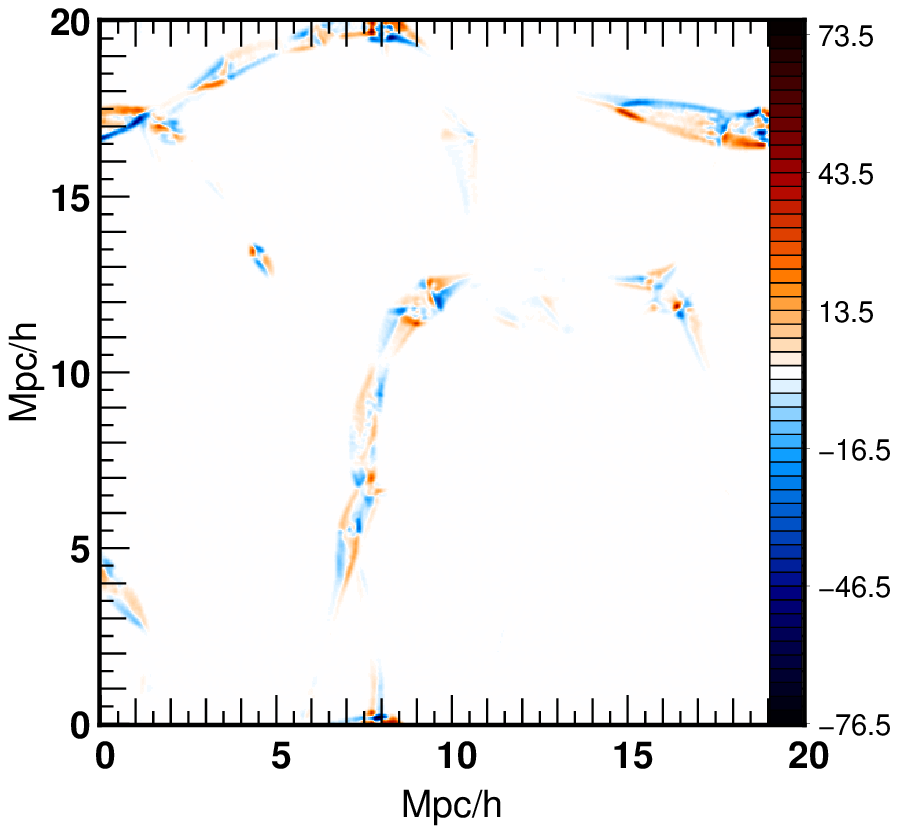}
\caption{A thin slice ($2 \, h^{-1}\, \rm Mpc$ thickness) of the projected DM density ({\sl left panels}) and the projected vorticity along the normal to the slice in unit of  $h\, \rm km \, s^{-1}\, Mpc^{-1}$ ({\sl right panels}). DM density is plotted with a logarithmic scale. Vorticity is computed after smoothing of the velocity field with a Gaussian filter of 160 $h^{-1}\, \rm kpc$ for this figure only. The geometry of the vorticity closely follows the LSS, but switches polarity across the walls/filaments (recalling that walls appear as filaments and filaments as peaks in such a cross-section). Note also how the vorticity is localized around filaments (the 2D peaks, as exemplified in Fig.~\ref{fig:rot-slice-25-white}). The two panels allow for a comparison between a section of ${\cal S}^{\rm CDM}_{20}$ ({\sl top}) and ${\cal S}^{\rm HDM}_{20}$ ({\sl bottom}). 
 In ${\cal S}^{\rm HDM}_{20}$, high-frequency modes are suppressed but the low-frequency vorticity is  qualitatively consistent with that found in  the  realistic ${\cal S}^{\rm CDM}_{20}$.
 On the bottom left panel, the density caustics are quite visible and correspond to the outer edge of the multi-flow region in the bottom right panel. }
\label{fig:CDMvsHDM}
\end{figure*}

All the statistical results of Section~\ref{sec:result} rely 
on a set of DM standard $\Lambda$cold dark matter ($\Lambda$CDM) simulations presented in Table \ref{tab:simu}. 
These simulations are characterized by the following $\Lambda$CDM cosmology: $\Omega_{\rm m}=0.24 $, $\Omega_{\Lambda}=0.76$, $n=0.958$, $H_0=73 \, \rm km. s^{-1} \, Mpc^{-1}$ and $\sigma _8=0.77$ within one standard deviation of 3 year Wilkinson Microwave Anisostropy Probe results \citep{Spergeletal03}. 

We use different box sizes: a 100 $h^{-1}\,  \rm Mpc$ box with an initial mean spatial resolution of 390 ${h}^{-1}\, \rm kpc$ ($256^3$ DM particles) in order to build a statistical sample of haloes and filaments, several 50 $h^{-1}\, \rm Mpc$ boxes with a mean spatial resolution of 190 ${h}^{-1}\, \rm kpc$ ($256^3$ particles), and a 20 $h^{-1}\, \rm Mpc$ box with a mean spatial resolution of 39 ${h}^{-1}\, \rm kpc$ ($512^3$ particles). All these simulations were run with {\sc gadget }\citep{GADGET}, using a softening length of $1/20^{\rm th}$ of the mean inter-particle distance.
 We also use the Horizon-$4\pi$ simulation, a 2000  $h^{-1}\,  \rm Mpc$ box  ${\cal S}^{\rm CDM}_{2000}$ with $4096^3$ DM particles \citep{Teyssier2009}. 
 
In addition, the 
$\Lambda$HDM subset corresponds to simulations with initial conditions that have been smoothed with a Gaussian filtering on scales of 2.3 and $0.23 h^{-1}\, \rm Mpc$, respectively, to suppress small-scale modes for the purpose of visualization and interpretation.  
All simulations but the sets ${\cal S}^{\rm CDM}_{50}$, ${\cal S}^{\rm CDM}_{2000}$ share the same phases.

All the simulations are studied at redshift $z=0$.
DM haloes are defined thanks to the Friend-of-Friend Algorithm  \citep[or FOF;][]{Huchra82}, with a linking length of $0.2 (L_{\rm box}^{3}/N_{\rm part})^{1/3}$.
 In the present work, we only consider haloes with more than 100 particles, which corresponds to a minimum halo mass of 62 $\times 10^{10}$ M$_{\odot}$ in ${\cal S}^{\rm CDM}_{50}$. 
The spin of a halo is defined as the sum over its particles $i$: $ \sum_{i} (\textbf{r}_{i}-\overline{\textbf{r}})\times (\textbf{v}_{i}-\overline{\textbf{v}})$ where $\overline{\mathbf{r}}$ is the centre of mass of the FOF and $\overline{\mathbf{ v}}$ its mean velocity. 
 As discussed in \cite{Pueblas:2008de}, for the DM simulations  we sample optimally the velocity field using  a Delaunay  tessellation. 
 
The FOF is prone to spuriously link neighbouring structures with tenuous bridges of particles, leading to artificial objects with a very high velocity dispersion, which could eventually bias the measure of the spin and consequently the alignment of the spin and the vorticity. 
Appendix~\ref{sec:Fof} investigates the effect of such spurious linkage on vorticity alignments and allow us to conclude that it does not impact the result.

The vorticity of the velocity is then measured from the resampled velocity at each point of the $256^{3}$ grid as the curl of the velocity field $\omega\,=\,\nabla \times \textbf{v}$, after Gaussian smoothing of the velocity field with a kernel length of 390 $h^{-1}\, \rm kpc$ for $ \displaystyle {\cal S}^{\rm CDM}_{100}$ and ${\cal S}^{\rm HDM}_{100}$, a kernel length of 780 $h^{-1}\, \rm kpc$ for $ \displaystyle {\cal S}^{\rm CDM}_{50}$ and a kernel length of 39 $h^{-1}\, \rm kpc$ for ${\cal S}^{\rm CDM}_{20}$ and ${\cal S}^{\rm HDM}_{20}$. The effect of the smoothing scale on the statistical alignments presented below is investigated in Appendix~\ref{sec:smoothing}. The results do not qualitatively depend on the smoothing scale and the main conclusion remains unchanged, even if the magnitude of the signal varies slightly (but not monotically) according to the scale.

A comparison between vorticity maps in ${\cal S}^{\rm CDM}_{20}$  and in ${\cal S}^{\rm HDM}_{20}$ is shown in Fig.~\ref{fig:CDMvsHDM}. 
Vorticity along the normal to the section is plotted in the right panels of this figure. In ${\cal S}^{\rm HDM}_{20}$, high frequencies features are suppressed but the low-frequency vorticity remains consistent with that of the more realistic ${\cal S}^{\rm CDM}_{20}$. 
In ${\cal S}^{\rm HDM}_{100}$, the smoothing is chosen such that
in high-vorticity regions (defined here as being regions where the vorticity is greater than 20 per cent of the maximum vorticity), the mean vorticity is of the order of 90 {$h$}\,$\rm km \, s^{-1}\, Mpc^{-1}$, i.e. it corresponds more or less to one revolution per Hubble time, in agreement with the theoretical predictions of \cite{pichon99}. The orders of magnitude are similar in ${\cal S}^{\rm CDM}_{100}$, ${\cal S}^{\rm HDM}_{20}$ and ${\cal S}^{\rm CDM}_{20}$.

The cosmic network is identified with {\sc rSeX} and {\sc DisPerSE}, the filament tracing algorithms based on either watersheding~\citep{sousbie09} or persistence~\citep{sousbie111,sousbie112} without significant difference for the purpose of this investigation.
The first method identifies ridges as the boundaries of walls which are themselves the boundaries of voids.
The second one identifies them as the `special' lines connecting topologically robust (filament-like) saddle points to peaks.
In this paper, the scale at which the filaments are traced (6 pixels Gaussian for each simulation) corresponds to large enough scales so that we are investigating the flow relative to the LSS
(though see Appendix~\ref{sec:persistence} for variations). Filaments are defined as a set of small segments linking neighbours pixels together. The mean size of the segments is 0.6 pixels, which means 234 $h^{-1}$ kpc in ${\cal S}^{\rm CDM}_{100}$.

For comparison with previous studies \citep[e.g.][]{libeskind13b}, walls are defined according to the density Hessian. Given $\lambda_{i}$ the eigenvalues of the Hessian ${ \cal H}={\partial^2\rho}/{\partial \text{r}_{i} \partial \text{r}_{j} }$ where $\rho$ is the density field, with $\lambda_{i} > \lambda_{j}$ if $i<j$,  walls are identified as being the region of space where $\lambda_{1} > \lambda_{2} >0$ and $\lambda_{3}<0$. The normal of a wall is given by the direction of the eigenvector associated with $\lambda_{3}$.
To obtain the Hessian, the density field of ${\cal S}^{\rm CDM}_{100}$ is smoothed with a Gaussian filter of 1.6 $h^{-1}$Mpc and differentiation of the density field is performed in  Fourier space.  

To estimate the number of multi-flow regions within the caustic and their size, for each segment of the skeleton, the vorticity cube is cut with a plane perpendicular to the direction of the filament. The number of multi-flow regions is given by the number of regions of positive and negative projected vorticity along this direction (with a given threshold), counted in a small window centred on the filament.  To obtain the size of the regions with a given polarity, the area where the absolute projected vorticity along the normal is greater than 10 per cent of the maximum vorticity is measured, and this area is divided by the number of quadrants. Assuming that these regions are quarter of discs, it yields the corresponding radius. This measure is done in ${\cal S}^{\rm HDM}_{100}$.

%%%%%%%%%%%
\section{Statistical alignments}\label{sec:result}
%%%%%%%%%%%
%%%%%%%%%%%
Let us first present robust statistical results derived from  sets of 
$\Lambda$CDM simulations and $\Lambda$HDM simulations for  comparison.

%%%%%%%%%%%
\subsection{Correlation between vorticity and filaments }
\label{sec:vorticity}

\begin{figure}
\includegraphics[scale=0.55]{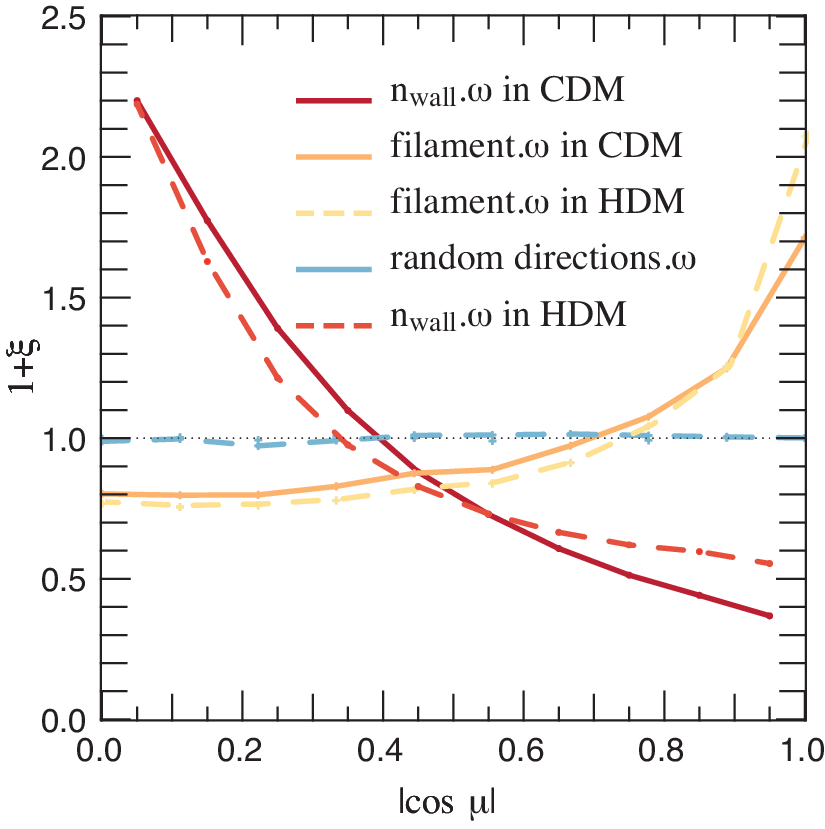} 
\caption{
The PDF of  $\cos \mu $, the cosine of the angle between the vorticity and the direction of the filament ({\sl orange}) and the angle between the vorticity and the normal of wall ({\sl red}) measured in the simulations ${\cal S}^{\rm CDM}_{100}$ (solid) and ${\cal S}^{\rm HDM}_{100}$ (dashed). The black dotted line corresponds to zero excess probability for reference. The large-scale total vorticity is preferentially aligned with  filament axis.}
\label{fig:cosinePDF}
\end{figure}

The alignment of vorticity with the direction of the filaments is  examined
in ${\cal S}^{\rm CDM}_{100}$ and in ${\cal S}^{\rm HDM}_{100}$.
The angle $\mu_{1}$ between the direction of the vorticity and the direction
of the filament is measured along each segment of the skeleton, and 
$\mu_{2}$ between the direction of the vorticity and the direction of the
normal of the wall. The probability distribution function (PDF)
of the absolute value of the cosine of these angles is shown in
Fig.~\ref{fig:cosinePDF}. This PDF is normalized for $\cos \mu$
between 0 and 1.  A strong detection is achieved. 
The signal is stronger in ${\cal S}^{\rm HDM}_{100}$
(because of a smoothing of high frequencies) but a clear signal
is also detected in ${\cal S}^{\rm CDM}_{100}$. 
As a check, the alignment between vorticity and shuffled segment directions
is then measured: no signal is detected.

In the filaments 
we find an excess of probability of 20 per cent to have
$\vert\cos\mu_{1} \vert$ in $\left[ 0.5,1\right]$
(that is $0\leq\mu_{1}\leq 60^{\rm o}$) relative to random orientations.
In the walls, we find an excess of probability of 45 per cent to have
$\vert\cos\mu_{2} \vert$ in $\left[ 0, 0.5\right]$ (that is
$60^{\rm o}\leq\mu_{2}\leq 90^{\rm o}$) relative to random orientations,
which means a strong signal for the vorticity to be aligned with the filament,
and perpendicular to the normal of the surrounding wall.

%%%%%%%%%%%%%%%%%%%%%
%%%%%%%%%%%%%%%%%%%%%
%
We conclude that in the neighbourhood of filaments, vorticity is preferentially aligned with the filament's axis and 
perpendicular to the normal of walls.
In other words, vorticity tends to be perpendicular to the axis along which material is collapsing fastest. This result is consistent with that of \cite{libeskind13b}, which explored the correlation between vorticity and shear eigenvectors. This correlation is confirmed in Appendix \ref{sec:validation}.

%%%%%%%%%%%%%%%%%%%%%%%%%
\subsection{Geometry of the multi-flow region}

\begin{figure*}
 \includegraphics[width=0.39\textwidth]{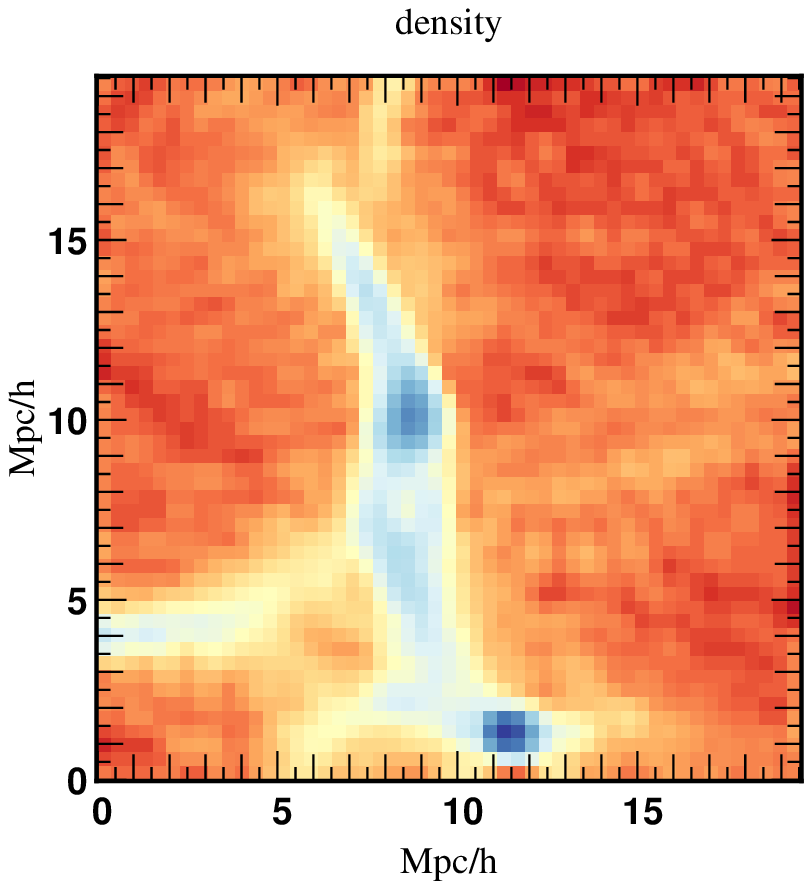}
 \includegraphics[width=0.45\textwidth]{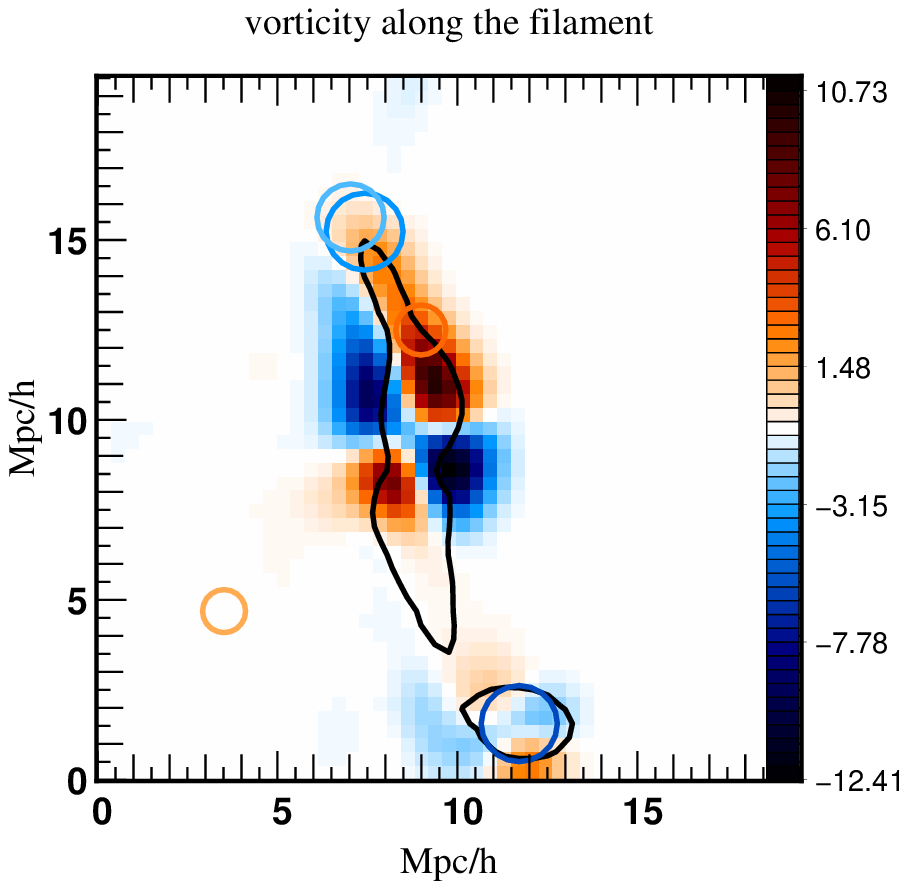}	
        \caption{Geometry/kinematics of a typical multi-flow region across a filament.{ \sl Left:} density map of a section
        perpendicular to a given filament in logarithmic scale.  {\sl Right}: projected vorticity along the filament within that section 
       (towards observer in red  and away from the observer in blue) in units of  $h\, \rm km \, s^{-1}\, Mpc^{-1}$ 
       on which is plotted in dark a contour of the density. Circles are haloes with their corresponding virial radius.
       The colour of the circles matches to the values of  $\cos \theta$ between the haloes spins and the normal
       of the section, positively oriented towards us. ${\cal S}^{\rm HDM}_{100}$ is used here, and for this figure only, vorticity is computed after smoothing the velocity field with a Gaussian filter of 1.6 $h^{-1}\, \rm Mpc$. 
        }\label{fig:vorticity}
\end{figure*}

\begin{figure}
\includegraphics[scale=0.9]{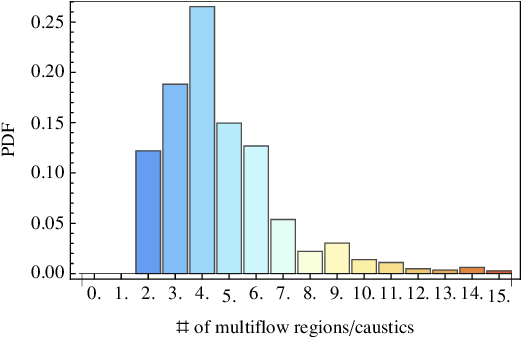} 
\vskip 0.2cm
\includegraphics[scale=0.9]{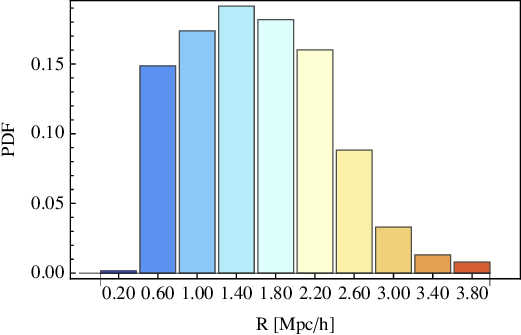} 
\caption{ {\sl Top:} normalized histogram of the number of multi-flow regions with different polarity around a filament measured in the simulation ${\cal S}^{\rm HDM}_{100}$. The mean corresponds to $\langle n_{\rm multiflow} \rangle =4.6$, the median is 4.25. On large scales, the multi-flow region is therefore typically quadrupolar. 
{\sl Bottom:} normalized histogram of the size of a  region in ${\cal S}^{\rm HDM}_{100}$ with a given polarity. The mean size of such region 
is  $\langle R \rangle= 1.6 \, h^{-1}\, \rm Mpc$, somewhat below the smoothing length of the initial conditions, $R_s=2.3 \, h^{-1}\, \rm Mpc$. 
It was checked on ${\cal S}^{\rm HDM}_{20}$that a similar scaling applies.
}
\label{fig:nbregion}
\end{figure}

Since section~\ref{sec:vorticity} showed that vorticity tends to be aligned with the filamentary features of the cosmic web, we are naturally led to focus on the structure of high-vorticity regions. The kinematics of the cross-sections of the filaments is therefore examined, by cutting our simulation with a plane perpendicular to the direction of the filament. We represent in this plane the projected vorticity along the filament.
Results are shown in Fig.~\ref{fig:vorticity} and can be summarized as follows: (i) vorticity is null outside the multi-flow region, and so confined to filaments (and walls in a weaker way) which is consistent with the assumption that cosmic flows are irrotational before shell-crossing; (ii) the cross-sections of the filaments are partitioned into typically quadripolar multi-flow regions (see Fig.~\ref{fig:nbregion}) where the vorticity is symmetric with respect to the centre of the (density) caustic  such that the global vorticity within that caustic is null (as expected);  the typical size of each quadrant is of the order of a smoothing scale (as shown in Fig.~\ref{fig:nbregion}); (iii) high-vorticity resides in the low-density regions of filaments:  vorticity is mainly located at the edge of the 
%%%%%%%%%%%%%
%%%%%%%%%%%%%
multi-flow region {\sl on} the caustic (see also Fig.~\ref{fig:radprofile}); vorticity is in fact typically null {\sl at} the peak of density. 
 (iv) Each quadrant of the multi-flow region is fed by multiple flows, originating from neighbouring walls (see Fig~\ref{fig:trace}).

%%%%%%%%%%%%%%%%%%%%%%%%%%%%%
\subsection{Correlation between vorticity and spin }

The  alignment of vorticity with filaments  on the one hand, and previous results about alignment (or orthogonality)
of the low-mass (high-mass) haloes spin with the filament and the shear eigenvectors \citep{codis12,libeskind12}
on the other hand, suggests to revisit the alignment of spin with the vorticity (previously examined by \citealp{libeskind12})
and to analyse in depth the correlation between vorticity and AM. 
The measurement is done by computing the
vorticity at the positions of the haloes and the projection, $\cos \theta$, between both normalized vectors. 
First note that haloes typically stand within one quadrant of the vorticity within filaments and not at the intersection of these quadrants, which is why the spin/vorticity alignment is strong.

%%%%%%%%%%%%%%%%%%%%%
%%%%%%%%%%%%%%%%%%%%%

\begin{figure}
\begin{center}
\includegraphics[scale=0.6]{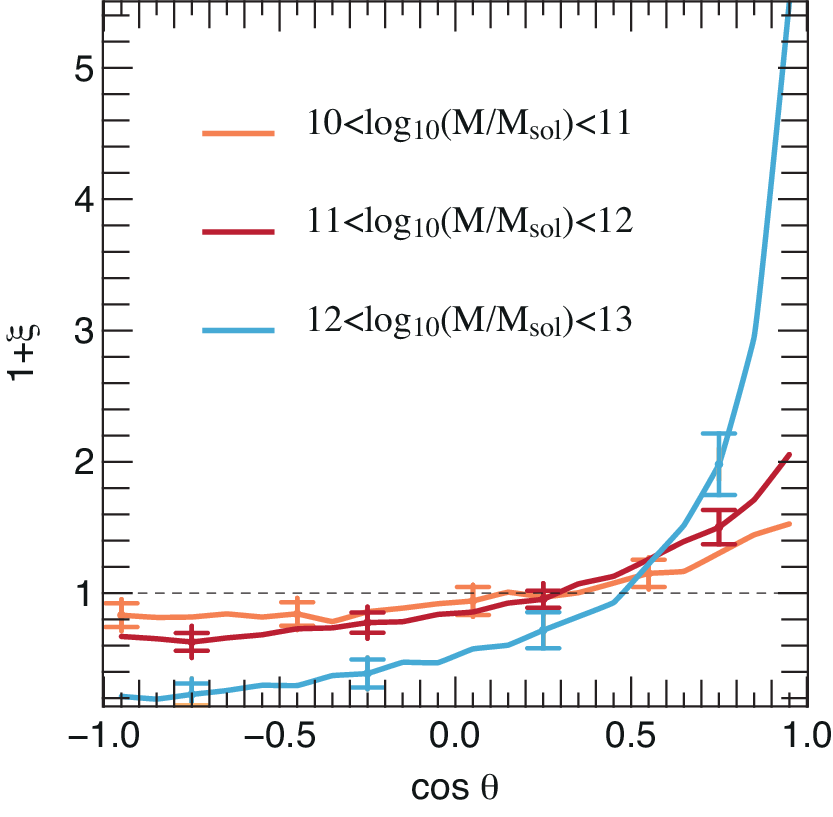}
\caption{The PDF of the angle between the vorticity
and the spin measured in 20 simulations of the  ${\cal S}^{\rm CDM}_{50}$ set.
Haloes are binned by mass as labeled.
The displayed error bars are 1-$\sigma$ standard deviation on the mean. 
 \label{fig:spin-multi}}
\end{center}
\end{figure}

The resulting PDF of  $\cos \theta$ is displayed in
Fig.~\ref{fig:spin-multi}. Here the set of simulations, ${\cal S}^{\rm CDM}_{50}$ are used to compute error bars on the correlation between spin and vorticity. 
The measured correlations are noisier as only a finite number of dark haloes are found within the simulation volume. 
It was checked that the correlation is not dominated by the intrinsic vorticity of the haloes themselves by computing the alignment between the spin and  the vorticity of the field {\sl after extruding} the FOF haloes, which led to no significant difference in the amplitude of the correlation.
We find an excess probability of 25 per cent relative to random orientations to have $\cos\theta$ in $\left[ 0.5,1\right]$ for haloes with $10\leq\log(\rm M/M_{\odot})\leq 11$, 55 per cent for $11\leq\log(\rm M/M_{\odot})\leq 12$ and 165 per cent for $12\leq\log(\rm M/M_{\odot})\leq 13$.
Note importantly that the intricate geometry of the multi-flow region (see also Figs.~\ref{fig:rot-slice-25-white} and \ref{fig:multisections}) strongly suggests retrospectively that the alignment (including polarity) between the spin of DM haloes and the vorticity of the flow within that region cannot be coincidental.

Fig.~\ref{fig:spin4pi}, which presents PDF of the cosine of the angle between the spin of 43 million dark haloes and the direction of the closest filament identified in the  ${\cal S}^{\rm CDM}_{2000}$ simulation, displays an interesting feature at low mass.
For the range of mass $\log M/\rm M_\odot \sim 11.5 - 12.5 $, the actual alignment between the spin and the direction of the filament {\sl increases} with mass,  before it becomes abruptly perpendicular around $5 \times10^{12} \, \rm M_\odot$.  This is fully consistent with the corresponding increase in vorticity  shown in Fig.~\ref{fig:spin-multi}, and will be discussed further in the next section.

%%%%%%%%%%%%%%%%%%%
\section{Interpretation}\label{sec:interpretation}
%%%%%%%%%%%%%%%%%%%
%%%%%%%%%%%%%%%%%%%

Let us now turn to the visualization of special purpose simulations, the $\Lambda$HDM set, to identify the origin and implications  of the measured vorticity of Section~\ref{sec:result},
and explain the observed mass transition.

%%%%%%%%%%%%%%%%%%%
\subsection{Building up vorticity from LSS flow}\label{sec:inter1}
%---------------------------------------------------

{Let us first show  that density walls are preferentially aligned with zero-vorticity walls}. 

Fig.~\ref{fig:rot-slice-25-white} displays the vorticity field in the neighbourhood of the main filament of the idealized `HDM' simulation, ${\cal S}^{\rm HDM}_{20}$. 
The vorticity bundle is clearly coherent on large scales, and aligned with the direction of the filament, strongest within its multi-flow core region,
while its  essentially quadrupolarity is twisted  around it. 

Fig.~\ref{fig:wall-in-vort} displays the cross-section of the vorticity perpendicular to the main filament shown in Fig.~\ref{fig:rot-slice-25-white}. 
The velocity field lines (in blue) converge towards the local walls (in brown) and are visually in agreement with 
the vorticity field which is partitioned by these walls. 
This picture is  qualitatively consistent 
with the scenario presented in \protect\cite{codis12}, as it shows that the filaments are fed via the embedding walls, while the geometry of the flow generates vorticity within 
their core. This vorticity defines the local environment  in which {DM haloes} form with a spin aligned with that vorticity.
The alignment between the contours of minimal vorticity and the density walls which is visually observed in Fig.~\ref{fig:wall-in-vort} (left panel) is then quantitatively examined. The probability distribution  of the cosine of the angle between the zero vorticity contour and the wall within the caustic is plotted on the right panel of Fig.~\ref{fig:wall-in-vort} (see Appendix~\ref{sec:walls} for the definition of the zero vorticity contour). An excess of probability of 15 per cent is observed for $\cos \psi $ in $\left[0.5,1 \right]$ relative to random distribution, that is for the alignment of the walls with the minimal vorticity contours.  This alignment increases with the smoothing of the tessellations, as expected.  
%%%%

 %%%%%%%%
 %%%%%%%%
\begin{figure}
\begin{center}
\includegraphics[scale=0.575]{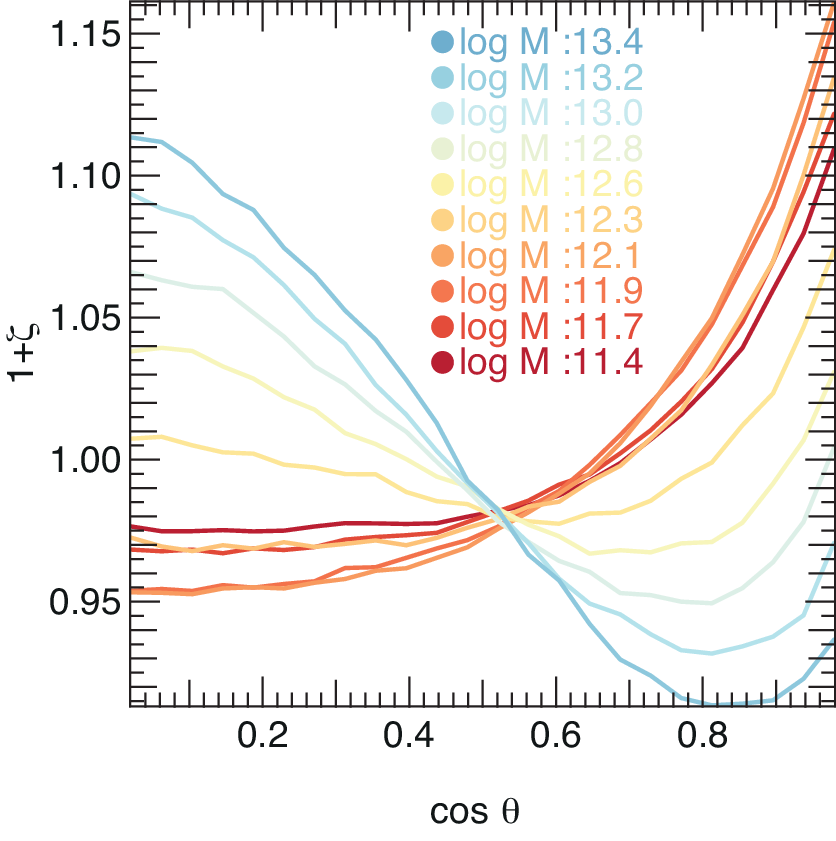}
\end{center}
\caption{The probability distribution of the cosine of the angle between the spin of dark haloes and the direction of the closest filament
as a function of mass in the ${\cal S}^{\rm CDM}_{2000}$ simulation. The smoothing length over which filaments are defined is 5 $h^{-1}\rm Mpc$. This figure extends the result first reported in \protect\cite{codis12}  to the mass range $\log \rm M/\rm M_{\odot}\sim$  11.5 - 12.0. In this mass range one observes that the probability to have  a small angle between the halo's spin and the filament's direction first{ \sl increases} (in red) as mass grows to $\log \rm M/\rm M_{\odot} \sim 12.1$, in agreement  with the increased spin--vorticity  alignment demonstrated in Fig.~\ref{fig:spin-multi}. At larger masses (from orange to blue) the statistical spin--filament alignment quickly decays, with a critical mass (in yellow) corresponding to a transition to predominately orthogonal orientations (in blue) at $\log \rm M_{\rm crit}/\rm M_\odot \approx 12.7$  as defined by \protect\cite{codis12}.
}
\label{fig:spin4pi}
\end{figure}
%%%%%%%%%

\subsection{Progenitors of multi-flow region}\label{sec:origin}
%---------------------------------------------------

In a DM (Lagrangian) simulation,  it is straightforward to identify the origin of particles within the multi-flow region.
Fig.~\ref{fig:trace} traces back in time DM particles ending up within a quadrant of the multi-flow region.
The quadrant is  fed by three flows of particles. The flow is  irrotational in the initial phase of  structure formation until the crossing of three flows in the vicinity of the filaments generates  shear and vorticity close to the caustic.

Note that the sharp rise near the  edge of the multi-flow region at the caustic
  is qualitatively consistent  with catastrophe theory \citep{Arnold92},  and is directly related to 
 the prediction of \cite{pichon99}.
Fig. \ref{fig:radprofile} illustrates this fact. To obtain this profile, a filament is cut in slices, corresponding to  filament segments: each slice corresponds to a plane perpendicular to the direction of the segment. Local vorticity is measured within that plane and stacked. The amount of vorticity is greater near the caustic. These  results are qualitatively consistent with the above-mentioned  theoretical predictions which characterize the size and shape of the multi-flow regions after  first shell crossing, and estimate their vorticity content as a function of cosmic time.

In short, having looked in detail at the set of (Lagrangian-) smoothed simulations allows us to conclude that 
  streaming  motion of DM {\sl away} from minima and wall-saddle points of the field,
and {\sl along} the walls of the density field is responsible for generating  the multi-flow region in
which vorticity arises. In turn, this vorticity defines the environment in which lower mass  haloes collapse.
Such haloes inherit  their spin from this environment, as quantified by Fig.~\ref{fig:spin-multi}.

\subsection{Mass transition for spin--filament alignment}\label{sec:mass}
%%%%%%%%%%%%%%

Up to now, we have not considered  the mass of the forming halo within the multi-flow region. The assumption has been that the Lagrangian extension of the progenitor of 
the dark halo was small compared to the antecedents of a given vorticity quadrant, so that the  collapse occurs within a quadrant of  a given  polarization, and leads to the 
formation of haloes with a spin parallel to that vorticity. 
For more massive objets (of the order of the transition mass), we can anticipate that their progenitor patch overlaps future vorticity quadrants of opposite polarity, hence that they will mostly cancel the component of 
their vorticity aligned with the filament as they form.The above-mentioned observed transition mass between aligned and anti-aligned spins relative to filaments 
would then typically correspond to the mass associated with the width of the quadrant of each caustics. 
In fact, as argued in \citet[fig~7]{pichon99} 
and shown on  Fig.~\ref{fig:radprofile}, 
the vorticity within the multi-flow region is mostly distributed near the caustic, 
on the  outer edge of the multi-flow region. It is therefore expected that, as the size of the collapsed halo increases, but remains {\sl below} that of the quadrant,
its vorticity should increase (as it collects more and more coherent rotating flow as secondary inflow), as shown in Fig.~\ref{fig:spin-multi}. As it reaches sizes {\sl above} that of the quadrant, it should start to
diminish significantly\footnote{In fact,  while investigating the statistics of the vorticity within spherical shells, \cite{pichon99} showed that if we consider spheres of size above 
one quadrant of the multi-flow regions, the total vorticity within that sphere drops significantly.} (see also Fig.~\ref{fig:caustic} and Appendix~\ref{sec:Toy} where this transition is illustrated with the help of a toy model).

%%%%%%%%%%%
\begin{figure*}
\begin{center}
\includegraphics[scale=0.25]{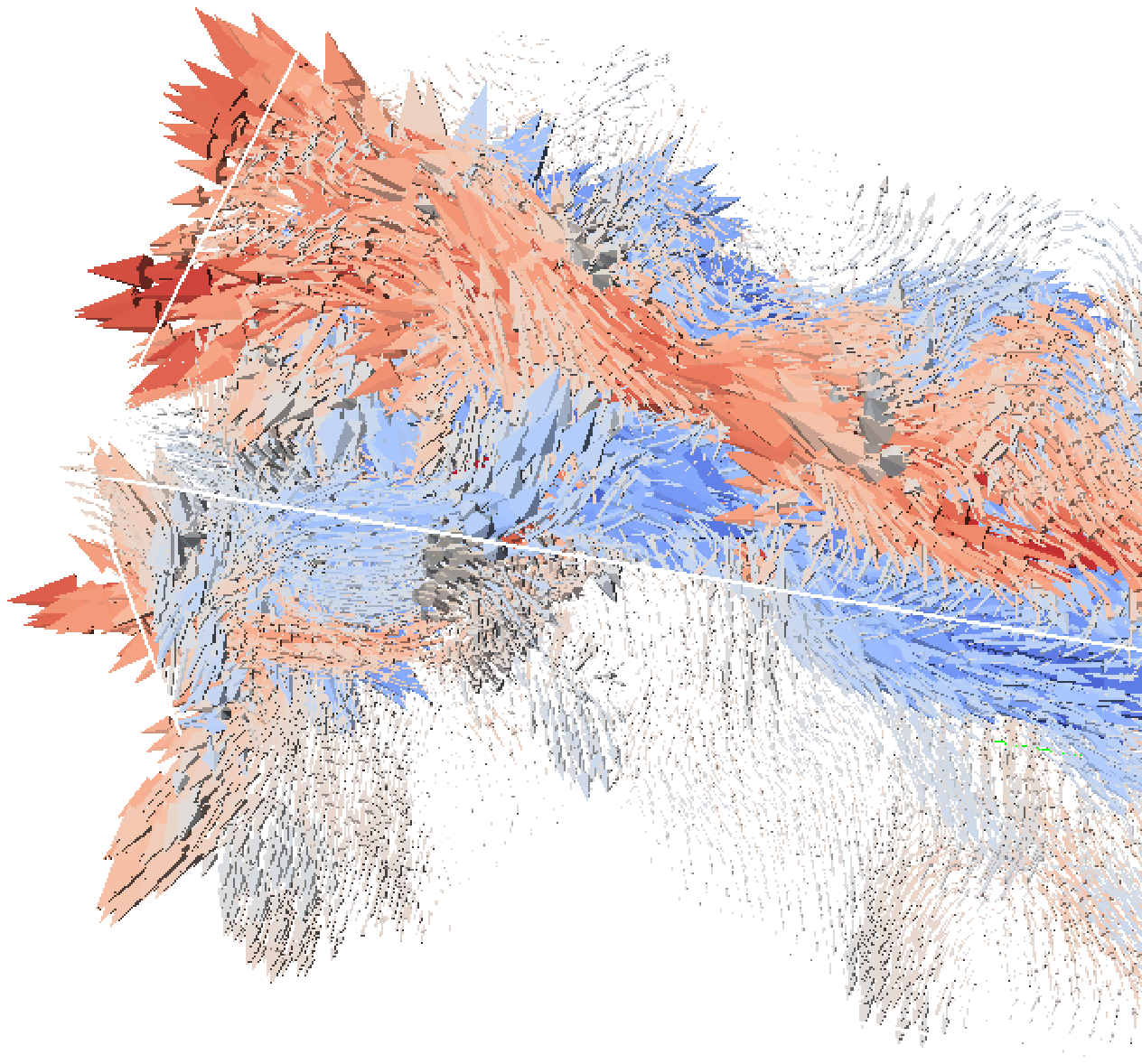} 
\end{center}
\caption{Vorticity field in the neighbourhood of the main filament of the idealized `HDM' simulation,
${\cal S}^{\rm HDM}_{20}$ 
 colour coded through its `z' component. 
The vorticity is clearly aligned with the direction of the filament, strongest within its multi-flow core region,
while its polarity is twisted around it. Helicity measurements are consistent with the observed  level of twisting.
We provide animations online at {\em\tt
http://www.iap.fr/users/pichon/spin/}.}
\label{fig:rot-slice-25-white}
\end{figure*}
%%%%%%%%%%%

Let us turn back specifically to Fig.~\ref{fig:spin4pi}.
For the range of mass $\log \rm M/\rm M_\odot\approx 11.4 - 12.1$,
the alignment between the spin and the direction of the filament
{\sl increases} with mass, peaking at $\rm M_{max} \approx 10^{12}\, M_\odot$,
before it rapidly decreases and changes to preferably 
perpendicular one for $\log \rm M > \log \rm M_{crit} \approx 12.7$, i.e 
$\rm M_{crit} \approx 5 \times 10^{12}  \, \rm M_\odot$.
This is fully consistent with the
corresponding increase in vorticity shown in Fig.~\ref{fig:spin-multi}.

The characteristic masses can be roughly understood by conjecturing that
the highest alignment occurs for the haloes which are of the size of vortices
in the caustic regions that \textit{just} undergo collapse. The measured
caustic structure depends on the chosen smoothing scale, so a recently formed
filament corresponds to the vortex
that shows a basic four quadrant structure and, following \cite{pichon99},
which  has vorticity close to the Hubble value $H$. 
From our simulations, the typical Lagrangian radius of such vortex
is  $\approx 1.5\, h^{-1} \, \rm Mpc$, which if taken  as the top-hat scale
gives a mass estimate $\rm M_{max} \approx 1.5 \times 10^{12}\, \rm M_\odot$
for the mass of haloes with maximum spin/filament alignment.
The transition to misalignment will happen at
$\rm M_{\rm crit} \approx 8 \times \rm M_{\rm max}$ where the whole width
of the filament is encompassed. Of course the quantitative accuracy of such
argument should not be over-emphasized. For instance, if we 
took the Lagrangian radius of the vortex to be $1.3 \, h^{-1}\, \rm Mpc$,
we would get $\rm M_{\rm max}=10^{12}\,  \rm M_\odot$, which would fit the transition of Fig.~\ref{fig:spin4pi} even closer.

%%%%%%%%%%%%%%%%%%%
\section{Discussions \& conclusions}\label{sec:conclusion}
%%%%%%%%%%%%%%%%%%%
%%%%%%%%%%%%%%%%%%%

Let us reframe the findings of Section~\ref{sec:result} and \ref{sec:interpretation} in the context of recent published results in this field, before concluding.

%%%%%%%%%%%%%%%%%%%
\subsection{Discussion}\label{sec:LSSvort}

%%%%%%%%%%%
\begin{figure*}
\includegraphics[width=0.49\textwidth]{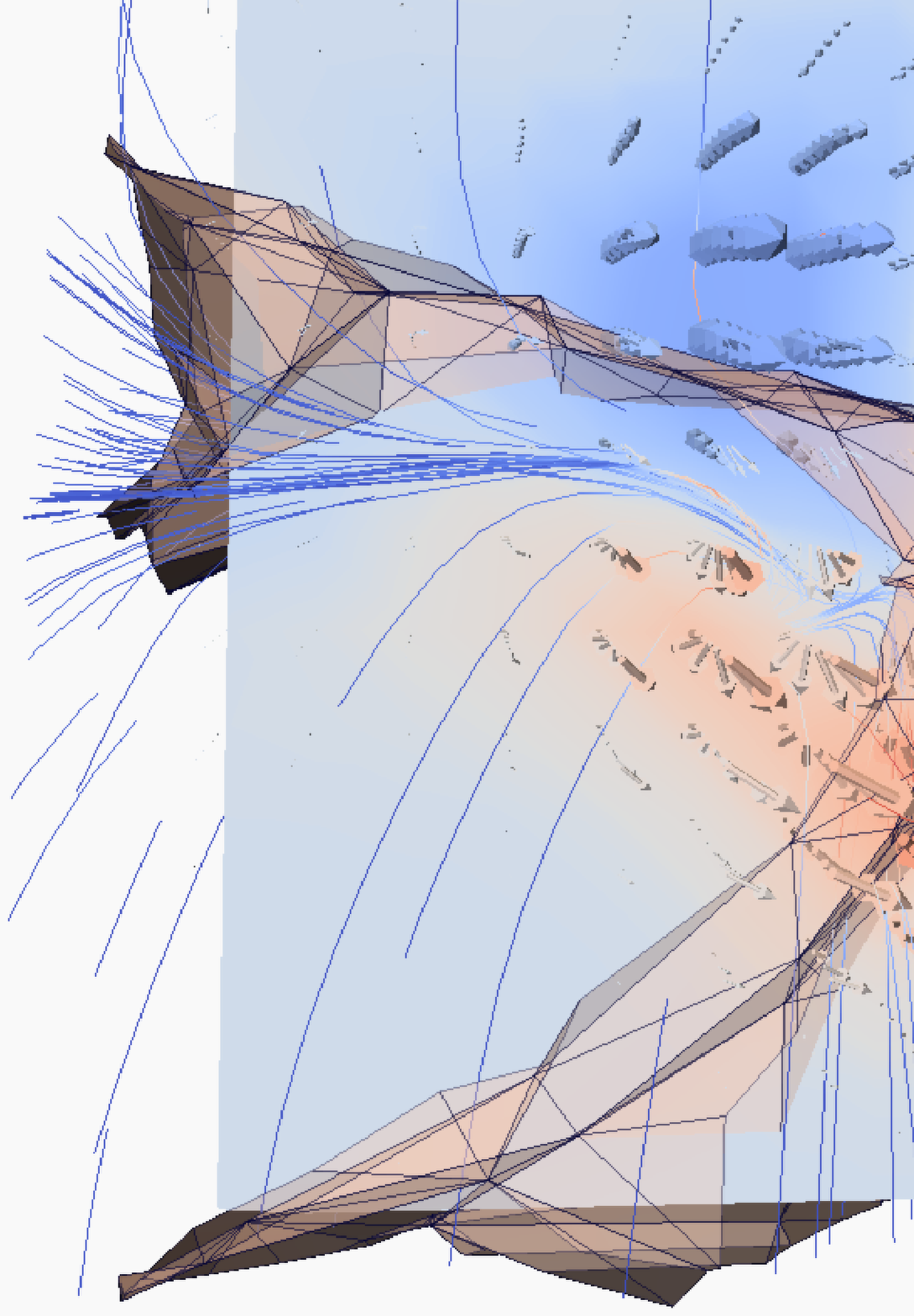} 
\includegraphics[width=0.37\textwidth]{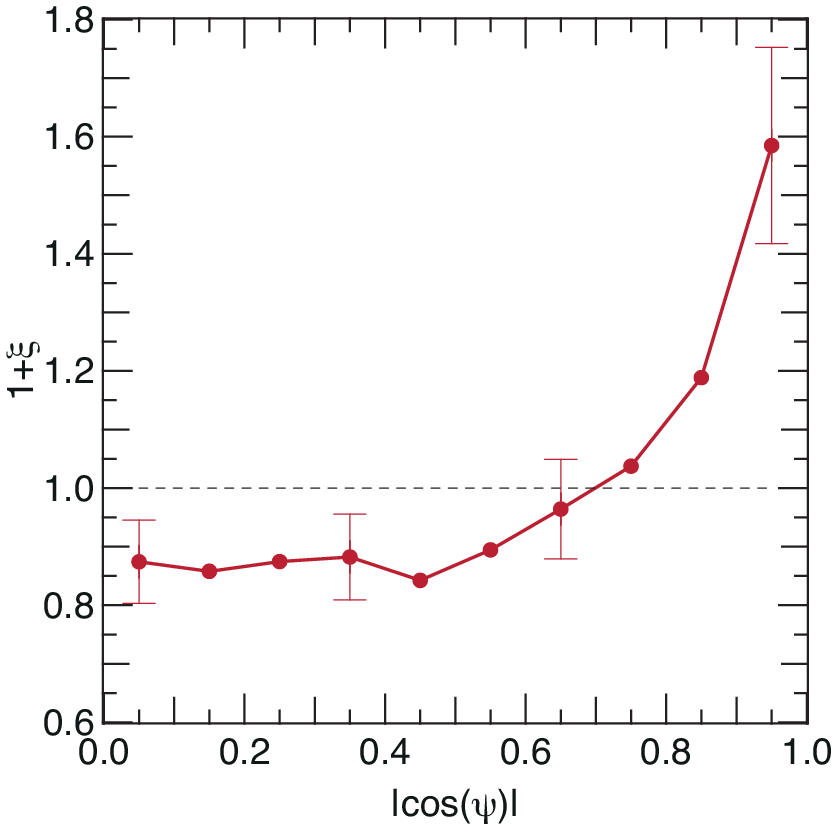} 
\caption{ {\sl Left:} the cross-section of the vorticity perpendicular to the main filament shown in Fig.~\ref{fig:rot-slice-25-white}. The colour coding in the section corresponds to the vorticity towards us (in blue) and away from us (in red) as shown by the corresponding arrows. The thin blue lines correspond to velocity field lines. The brown surfaces
represent the local walls.  The field lines converge towards the local walls and are in agreement with 
the vorticity field which is partitioned by these walls. 
{\sl Right:} the probability distribution as a function of the cosine angle between the normal to the zero vorticity walls and the normal to the density walls, $\cos \psi$, computed on the simulation ${\cal S}^{\rm CDM}_{100}$. The simulation is divided into eight 50 $h^{-1}\, \rm Mpc$ sub-boxes. Density walls are computed  using {\sc DisPerSE}, and the smoothing coefficient of the tessellation is S=4 (see Appendix~\ref{sec:walls}). The plotted signal corresponds to the average of the PDFs for the eight sub-boxes. The displayed error bars are 1$\sigma$ standard deviation on the mean. }
\label{fig:wall-in-vort}
\label{fig:zerovort}
\end{figure*}

\cite{libeskind13b}'s description of AM acquisition occurs in two stages (first through TTT, and then through the curl of the embedding velocity field). Results of sections~\ref{sec:result} and \ref{sec:interpretation} seem consistent with this. 
In particular the alignment of vorticity with the eigen vectors of the tidal field is confirmed in Appendix~\ref{sec:validation}. 
The  connection between \cite{pichon11} and this paper is  the following: in the former, it was shown that the spin up of dark haloes proceeded in stages: a given collapsing 
halo would first acquire some {specific angular} momentum following TTT, at turn around freezing its amplitude at the TTT expected value;
in a secondary stage (see their fig.~9), it would 
spin up again as it acquires {specific AM} from secondary infall coming from the  larger scale distribution of matter
which collapses at the next stage of hierarchical clustering. For relatively isolated massive haloes that form from statistically
rare density enhancements as studied in \cite{pichon11}, this secondary collapse just leads to a virialized halo of  increased mass.
The (Eulerian) emphasis of the current paper and of \cite{codis12} \citep[see also][]{danovichetal11} is to note that
for less massive and less rare haloes, 
forming in large-scale filamentary regions,
this secondary infall, coming late from the turn-around of the encompassing filamentary structure, is  arriving along marked preferred directions and is typically multi-flow and vorticity rich. 
Given that the shell crossing occurring during the later formation of that embedding filament generates vorticity
predominantly aligned with the filament, this secondary infall will contribute extra spin-up  along the 
filament direction.
Hence the global geometry of the inflow  is consistent with a spin acquisition for  haloes induced by the large-scale
dynamics within the cosmic web, and in particular its multi flow  vortices. 
This scenario may only be reconciled with the standard (Lagrangian) tidal torque theory if the latter
is augmented so as to account for the larger scale anisotropic environment of walls
 and filaments responsible for secondary infall \citep[see][]{pichon2014}.

%One can anticipate that taking into account the anisotropy of the longwave modes 
%that form the environment of the proto-halo in the spirit of peak-background split formalism \citep[][PBS]{kaiser84}
%will allow us to explain the alignment
%of the haloes' spin   within filaments. 
%Indeed, it can be shown that the presence of  a  filament {\sl and } a wall will impact the possible mis-alignment of the tidal tensor  and the inertia tensor of a collapsing peak.
% On the basis of what was found in this paper,  
%the following extra caveats to the extension of  PBS should nonetheless also be added for low mass dark haloes (below the transition mass): i)
%the DM particles originating from its  original Lagrangian patches reach the  filament from multiple directions;
%ii)  the flow from the first two Lagrangian patches (see Fig.~\ref{fig:trace}) reaching the same Eulerian
%patch within that filament may tidally spin-up  the proto-halo (embedded in the third patch) in a manner 
%which is not natively taken into account by (unconstrained) TTT. 
% In short, single shell unconstrained   TTT  describes well the process of AM acquisition for a halo up to the point when
%non-linearities on scales  larger than that of the halo  grow and turn around, as they may spin up additionally the halo consistently with the shape of the anisotropic  environment. 
%Depending on the rareness of the collapsing object, the two stages (primary versus secondary infall) may or 
%may not be distinct and overlap in time.

%
%%%%%%%%%%%%%
\begin{figure}
\includegraphics[width=0.49\textwidth]{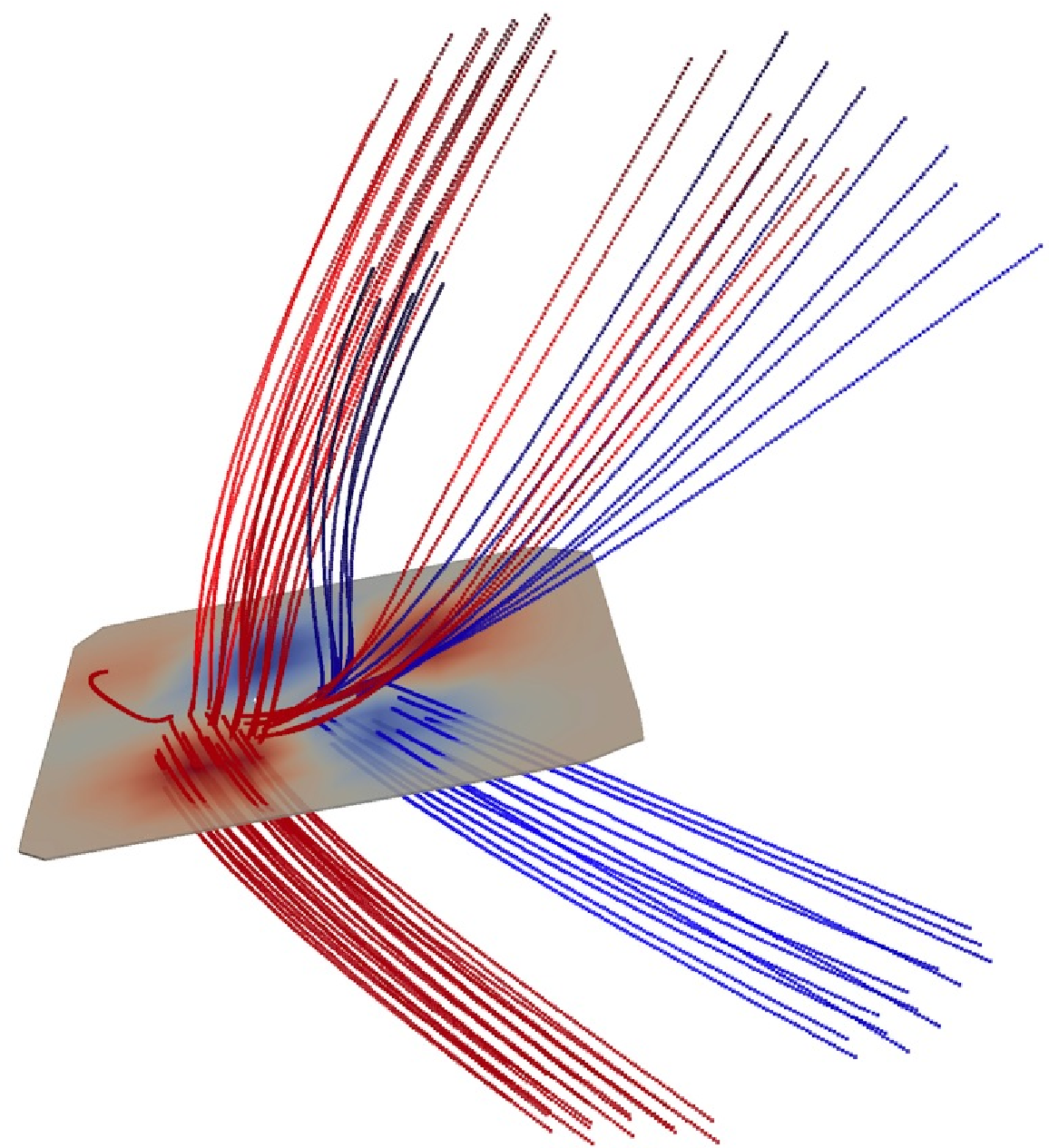}
\caption{{\sl Left}: 
individual DM particle trajectories ending in a given quadrant of the vorticity  multi-flow region. In blue are  particles ending in a region of positive projected vorticity along the filament, and in red are the particles ending in the negative  vorticity region. 
The quadrant is  fed by at least three flows of particles (see also the inset in Fig.~\ref{fig:caustic}, which represents qualitatively the theoretical expectation of
the starting points of these bundles in the Zel'dovich approximation.). 
The  ${\cal S}^{\rm HDM}_{100}$ simulation is used for this figure. }\label{fig:trace}
\end{figure}
%%%%%%%%

Let us sketch the basis of such calculation. The geometry of the setting is shown in Fig.~\ref{fig:caustic}.
For Gaussian random fields,
 one can  compute  the most likely tidal field and inertia tensor  at a given Lagrangian peak, subject to a Zel'dovich boost
which will translate that peak near  to a filament  at some distance $\mathbf{r}$; this distance corresponds to the time during which the nearby filament  has shell-crossed
 multiplied by the original velocity. In turn, the condition of shell crossing can be expressed as constraints on 
 the eigenvalues of the shear tensor. 
We can anticipate that the pre-existing Lagrangian
 correlation between the tidal field of the halo-to-be on the one hand, and the Hessian of the filament-to-be on the other hand,
  imposes  some  alignment between the direction of the filament (along the first eigenvector of the Hessian)
and the spin of the collapsing halo (as set by the corresponding tidal tensor).
If the critical condition that the filament is embedded into  a given wall is added, the axial symmetry of the problem
 will be broken, and the inertia and tidal tensor (which are sensitive to different scales) 
 will end up misaligned, reflecting this anisotropy.  
In this context, the observed spin (and importantly its polarity)  will  correlate with   the polarity of the vorticity quadrant the halo ends up into after translation.
The upshot is that in Fig.~\ref{fig:caustic}; the lighter pink sphere will `know' about the green dots given these constraints.
This supplementary requirement is imposed by the fact that the correlation between spin and vorticity keeps tract of the direction of both vectors, as shown in Fig.~\ref{fig:spin-multi}.
It appears from this sketch that, as the Lagrangian patch of the proto-halo  becomes of the order of  the typical Lagrangian size of the
 quadrant, the alignment will increase, and as it becomes larger, it will fade (see Appendix~\ref{sec:Toy} for an illustration of this transition).
 Note finally that this `one slice perpendicular to the filament axis' picture cannot address the process of  spin flipping to a
  perpendicular direction to the filament via mergers, as this is a longitudinal process. 
This is  also the topic of  \cite[][]{pichon2014} which complements the Eulerian view  presented here.
%%%%%%%
\begin{figure}
\begin{center}
\includegraphics[scale=0.5]{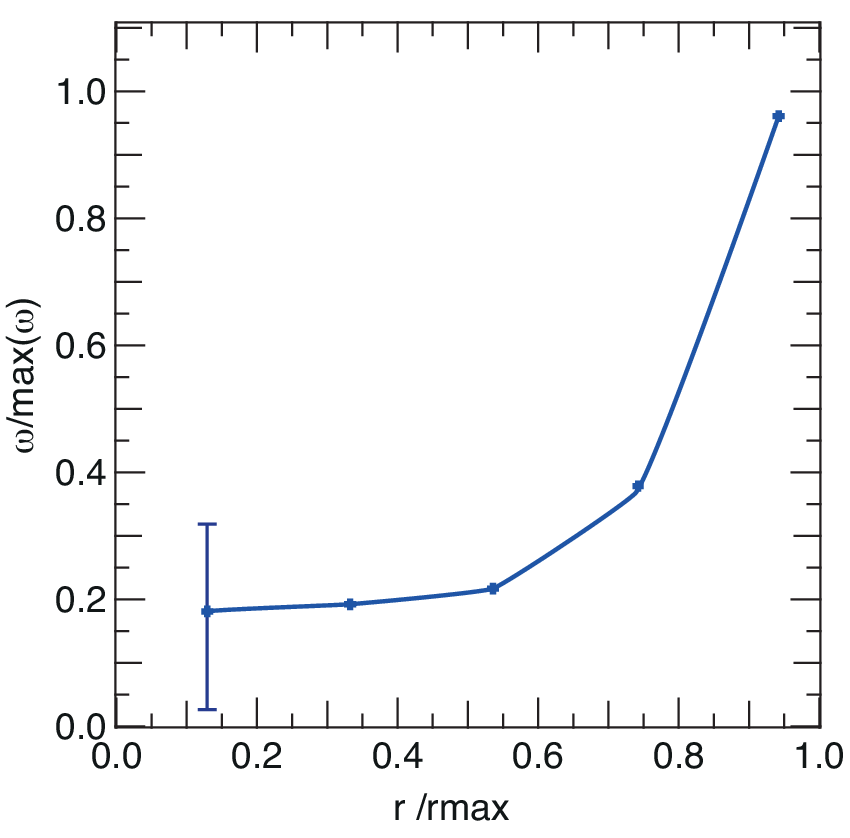}
\end{center}
\caption{Azimuthal average of the radial profile of the vorticity. The profile is obtained by averaging on the sections of a complete filament (each section is associated to a filament segment, to which the section is perpendicular). 
Vorticity is clearly larger towards the caustic, and would  theoretically  become singular (as $1/\sqrt{ 1- r/r_{\rm max}}$)  {\sl at } the caustic for a Zel'dovich mapping, as shown in \protect\cite{pichon99}.
Here the profile is convolved by shape variations from one caustic to another and by the azimuthal average.
The indicative error bar was computed  as the average over a larger  stack. }
\label{fig:radprofile}
\end{figure}
%%%%%%%%

%%%%%%%%%%%%%%%%%%%
\subsection{Conclusions}

\begin{table}
\singlespacing
\begin{tabular}{| l c l c | c}
 \textbf{Definition} & \textbf{Name} & \textbf{     Mean      } & \textbf{     Median      }\\ \hline
 \multicolumn{4}{c |}{ Alignment between vorticity and Cosmic Web} \\
  DM: vorticity/filaments & \multirow{3}{*}{$ |\cos \mu |$} &0.58 (0.5) & 0.62  \\
  Hydro: vorticity/filaments & & 0.58 (0.5) & 0.63 \\
  DM: vorticity/walls  &  & 0.34 (0.5) & 0.27 \\
  \hline
    \multicolumn{4}{c |}{Alignment between vorticity and haloes spin} \\
 $10 \leq  \log( M/ \rm M_{\odot}) \leq 11$ &   \multirow{3}{*}{$ \cos \theta$}  & 0.09 (0.0) & 0.14 \\
 $10 \leq  \log( M/ \rm M_{\odot}) \leq 12$ &    & 0.19 (0.0) & 0.29 \\
 $12 \leq  \log( M/ \rm M_{\odot}) \leq 13$ &    & 0.53 (0.0) & 0.72 \\
 \hline
   \multicolumn{4}{c |}{ Alignment between density walls and 0-vorticity walls} \\
  & $|\cos \psi |$ &  0.54 (0.5) & 0.56  \\
  \hline
  \multicolumn{4}{c |}{ Alignment between vorticity and tidal tensor eigenvectors} \\
vorticity / e1 &  \multirow{3}{*}{$ |\cos \gamma|$}& 0.62 (0.5)  & 0.69 \\
vorticity / e2 &  & 0.48 (0.5) & 0.47 \\
vorticity / e3 & & 0.31 (0.5) & 0.23 \\
\hline
\end{tabular}
\caption{The median and mean cosine values for the set of studied alignments. In parenthesis are the expected values for random distributions.}
\label{tab:mean}
\end{table}

Using large-scale cosmological simulations of structure formation, we have analysed the kinematic properties of the velocity flows relative to the cosmic web.
Our findings are the following.
\begin{itemize}
\item The vorticity in large-scale structures on scales of 0.39 $h^{-1}\, \rm Mpc$ and above is confined to, and aligned with, its filaments with an excess of probability of 20 per cent relative to random orientations, and perpendicular to the normal of the dominant walls at a similar level. 
This is consistent with the corresponding direction of the eigenvectors of the tidal field (and is expected given that the potential is a smoothed version of the density field).
\item At these scales, the cross-sections of these filaments are typically partitioned  into quadripolar caustics, with opposite vorticity parallel to their filament,
 arising from multiple flows originating from  neighbouring walls,  as would  secondary shell crossing  along these walls imply. 
 The radial vorticity profile within the multi-flow region displays a sharp rise near the caustic, a qualitatively expected feature  of catastrophe theory.
 \item The spins of embedded haloes within these filaments are consistently aligned with the vorticity of their host vorticity quadrant at a level of 165 per cent. The progenitor of lighter haloes within the multi-flow region can be traced back to three flows or more originating  from the neighbouring walls, and form within the filament.
 \item Appendix A shows that for adiabatic/cooling hydrodynamical simulations within the DM caustics,   the gas and the DM share the same vorticity orientation on large scales.  High-resolution cooling runs show that the small-scale structure of the velocity flow around forming galaxies does not destroy this larger scale coherence. 

\item The mass transition for spin--filament alignment is set  by  the size of sub-caustics with a given polarization (see Appendix~\ref{sec:Toy}). The alignment is strongest for Lagrangian patch commensurable with the sub-caustic as vorticity is strongest on the edge of the multi-flow region. Once the collapsed halo has a size larger than any such sub-caustic, it cancels out most of the vorticity within
  the caustics. 
 \end{itemize}

%%%%%%%
\begin{figure*}

\noindent\makebox[2.15\textwidth]{\includegraphics[scale=0.45]{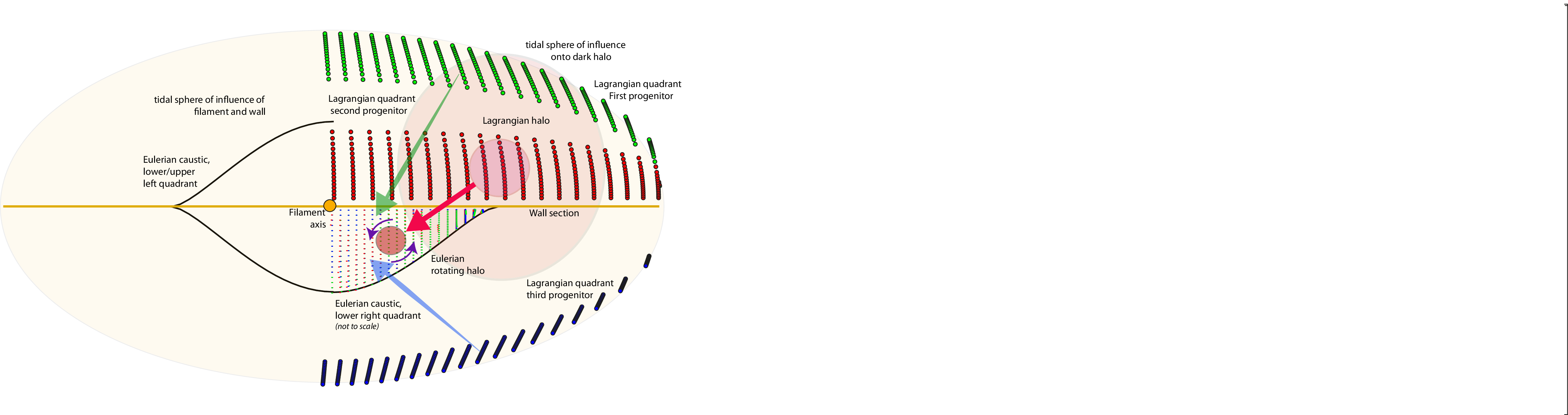}}

\caption{Sketch of the dynamics of a low mass halo formation and spin-up within a wall near a filament, which   are perpendicular to the plane of the image ({\sl in yellow}). The tidal sphere of influence of this structure is represented by the pale yellow ellipse. The three bundles of  large dots ({\sl in green, red, and blue}) represent Lagrangian points (at high redshift) which
image, after shell crossing,  will end up sampling regularly the lower right quadrant of the Eulerian multi-flow region; the three progenitor bundles
 are computed here in the Zel'dovich approximation \protect\cite[see][for details; note that this Eulerian quadrant is not up to scale]{pichon99}. 
Each pair of dots (one large, one small) represents the same DM particle in the initial condition and final condition.
In black, 3/4 of the Eulerian caustic.
In light, resp. dark {\sl pink}, the locus of the Lagrangian and Eulerian position of the halo, which has moved by 
a distance displayed by the {\sl red } arrow, and spun up following the purple arrows  while 
entering the quadrant. 
The {\sl blue} and {\sl green} arrows represent the path of fly-by DM particles originating from the other 
two bundles, which will contribute to torquing up the halo  \protect\cite[following][]{codis12}. Given the geometry of the flow imposed by 
the wall and swirling filament, the spin of the DM halo will necessarily be parallel to the direction of the filament
{\sl and} to the vorticity in that quadrant. 
In the language of TTT, the tidal field imposed on to the Lagrangian patch of the halo 
({\sl very light pink},  corresponding to secondary infall) should 
be evaluated subject to the constraint that the halo will {\sl move} into the {\sl anisotropic} {\sl multi-flow} region
(each emphasis imposing a constraint of its own);
these constraints will in turn impose that the corresponding spin-up will be aligned with the vortex. 
 }
\label{fig:caustic}
\end{figure*}
%%%%%%%%

The focus of  this paper was in explaining  the `where':   pinning down the locus of vorticity and describing the geometry of multi-flow infall towards filaments; 
and the `how':   explaining its origin via shell crossing. It  also provided an explanation for the origin of the 
mass transition for spin alignment. 
All measured  alignments are summarized in Table~\ref{tab:mean}.

Improvements beyond the scope of this paper include 
(i) developing the sketched  anisotropic (filamentary) peak-background-split theory of spin acquisition;
(ii) quantifying the curvilinear evolution of the vorticity (orientation and amplitude) as a function of distance to the critical points of the cosmic web and predicting the spin flip for high
masses;
(iii) quantifying the helicoidal nature of gas infall on galactic scales;
(iv) connecting the findings of this paper to the actual process of {\sl galactic}  alignment. 
\par
In turn, this should allow astronomers to shed light on the following problems:
how and when  was the present Hubble sequence of galaxies established?
How much of the dynamical evolution of galaxies is driven by environment?
What physical processes transforming galaxies dominate morphology: 
galaxy interactions and mergers, external accretion and outflows, secular evolution?
What is their respective roles in shaping discs, bulges or spheroids? 
Is it the same process at low and high redshift?
These are  addressed in part  in the companion paper, \cite{Dubois2014}, which shows  in particular using state-of-the-art hydrodynamical simulations with AGN/SN feedback that at high redshifts the large vorticity of the gas flow is correlated with the direction of the spin of {\sl galaxies} (their fig. 12).

\subsection*{Acknowledgments}
We thank  M. Haehnelt, James Binney and D. Lynden-Bell  for useful comments during the course of this work. 
JD and AS's research is supported by Adrian Beecroft, the Oxford Martin School and STFC.
Let us thank D.~Munro for freely distributing his {\sc \small  Yorick} programming language and {\sc \small opengl} interface (available at {\tt http://yorick.sourceforge.net}).
Some visualizations made use of the software {\sc \small paraview} {\tt http://www.paraview.org}.
This work is partially supported by the Spin(e) grants ANR-13-BS05-0005 of the French {\sl Agence Nationale de la Recherche}
and by the ILP LABEX (under reference ANR-10-LABX-63 and ANR-11-IDEX-0004-02).
CP thanks the PEPS  `Physique th\'{e}orique et ses interfaces' for funding,
the institute of Astronomy for a Sacker visiting fellowship and  the KITP for hospitality via  the National Science Foundation under Grant No. NSF PHY11-25915. 

\bibliographystyle{mn2e}
\bibliography{refs}

\begin{thebibliography}{}

\bibitem[\protect\citeauthoryear{Arag\'on-Calvo, van~de Weygaert, Jones \&
  van~der Hulst}{Arag\'on-Calvo et~al.}{2007}]{Calvo07}
Arag\'on-Calvo M.~A.,  van~de Weygaert R.,  Jones B. J.~T.,    van~der Hulst
  J.~M.,  2007, \apj, 655

\bibitem[\protect\citeauthoryear{{Aragon-Calvo} \& {Yang}}{{Aragon-Calvo} \&
  {Yang}}{2014}]{Calvo13}
{Aragon-Calvo} M.~A.,  {Yang} L.~F.,  2014, \mnras, 440, L46

\bibitem[\protect\citeauthoryear{Arnold}{Arnold}{1992}]{Arnold92}
Arnold V.~I.,  1992, Catastrophe theory / V.I. Arnold ; translated from the
  Russian by G.S. Wassermann ; based on a translation by R.K. Thomas, 3rd rev.
  and expanded ed. edn.
Springer-Verlag Berlin ; New York

\bibitem[\protect\citeauthoryear{{Aubert}, {Pichon} \& {Colombi}}{{Aubert}
  et~al.}{2004}]{aubert04}
{Aubert} D.,  {Pichon} C.,    {Colombi} S.,  2004, \mnras, 352, 376

\bibitem[\protect\citeauthoryear{Bailin \& Steinmetz}{Bailin \&
  Steinmetz}{2005}]{b24}
Bailin J.,  Steinmetz M.,  2005, \apj, 627

\bibitem[\protect\citeauthoryear{Bett, Eke, Frenk, Jenkins, Helly \&
  Navarro}{Bett et~al.}{2007}]{Bett21032007}
Bett P.,  Eke V.,  Frenk C.~S.,  Jenkins A.,  Helly J.,    Navarro J.,  2007,
  Monthly Notices of the Royal Astronomical Society, 376, 215

\bibitem[\protect\citeauthoryear{{Birnboim} \& {Dekel}}{{Birnboim} \&
  {Dekel}}{2003}]{birnboim03}
{Birnboim} Y.,  {Dekel} A.,  2003, \mnras, 345, 349

\bibitem[\protect\citeauthoryear{{Codis}, {Pichon}, {Devriendt}, {Slyz},
  {Pogosyan}, {Dubois} \& {Sousbie}}{{Codis} et~al.}{2012}]{codis12}
{Codis} S.,  {Pichon} C.,  {Devriendt} J.,  {Slyz} A.,  {Pogosyan} D.,
  {Dubois} Y.,    {Sousbie} T.,  2012, \mnras, 427, 3320

\bibitem[\protect\citeauthoryear{{Danovich}, {Dekel}, {Hahn} \&
  {Teyssier}}{{Danovich} et~al.}{2012}]{danovichetal11}
{Danovich} M.,  {Dekel} A.,  {Hahn} O.,    {Teyssier} R.,  2012, \mnras, 422,
  1732

\bibitem[\protect\citeauthoryear{Doroshkevich}{Doroshkevich}{1970}]{TTTR}
Doroshkevich A.~G.,  1970, Astrofizika, 6, 581

\bibitem[\protect\citeauthoryear{{Dubois}, {Pichon}, {Haehnelt}, {Kimm},
  {Slyz}, {Devriendt} \& {Pogosyan}}{{Dubois} et~al.}{2012}]{Dubois2011}
{Dubois} Y.,  {Pichon} C.,  {Haehnelt} M.,  {Kimm} T.,  {Slyz} A.,  {Devriendt}
  J.,    {Pogosyan} D.,  2012, \mnras, 423, 3616

\bibitem[\protect\citeauthoryear{{Dubois}, {Pichon}, {Welker}, {Le Borgne},
  {Devriendt}, {Laigle}, {Codis} \& {Pogosyan}}{{Dubois}
  et~al.}{2014}]{Dubois2014}
{Dubois} Y.,  {Pichon} C.,  {Welker} C.,  {Le Borgne} D.,  {Devriendt} J.,
  {Laigle} C.,  {Codis} S.,    {Pogosyan} D.,  2014, \mnras, 444, 1453

\bibitem[\protect\citeauthoryear{{Haardt} \& {Madau}}{{Haardt} \&
  {Madau}}{1996}]{haardt&madau96}
{Haardt} F.,  {Madau} P.,  1996, \apj, 461, 20

\bibitem[\protect\citeauthoryear{{Hahn}, {Carollo}, {Porciani} \&
  {Dekel}}{{Hahn} et~al.}{2007}]{hahn07}
{Hahn} O.,  {Carollo} C.~M.,  {Porciani} C.,    {Dekel} A.,  2007, \mnras, 381,
  41

\bibitem[\protect\citeauthoryear{Hoyle}{Hoyle}{1949}]{Hoyle49}
Hoyle F.,  1949, Problems of Cosmical Aerodynamics, Central Air Documents,
  Office, Dayton, OH.
Central Air Documents Office, Dayton, OH

\bibitem[\protect\citeauthoryear{Huchra \& Geller}{Huchra \&
  Geller}{1982}]{Huchra82}
Huchra J.~P.,  Geller M.~J.,  1982, \apj, 257, 423

\bibitem[\protect\citeauthoryear{{Katz}, {Keres}, {Dave} \& {Weinberg}}{{Katz}
  et~al.}{2003}]{katz03}
{Katz} N.,  {Keres} D.,  {Dave} R.,    {Weinberg} D.~H.,  2003, in
  {J.~L.~Rosenberg \& M.~E.~Putman} ed., The IGM/Galaxy Connection. The
  Distribution of Baryons at z=0 Vol.~281 of Astrophysics and Space Science
  Library, {How Do Galaxies Get Their Gas?}.
pp 185--191

\bibitem[\protect\citeauthoryear{{Kere{\v s}}, {Katz}, {Weinberg} \&
  {Dav{\'e}}}{{Kere{\v s}} et~al.}{2005}]{keresetal05}
{Kere{\v s}} D.,  {Katz} N.,  {Weinberg} D.~H.,    {Dav{\'e}} R.,  2005,
  \mnras, 363, 2

\bibitem[\protect\citeauthoryear{{Kimm}, {Devriendt}, {Slyz}, {Pichon},
  {Kassin} \& {Dubois}}{{Kimm} et~al.}{2011}]{kimm11}
{Kimm} T.,  {Devriendt} J.,  {Slyz} A.,  {Pichon} C.,  {Kassin} S.~A.,
  {Dubois} Y.,  2011, ArXiv:1106.0538

\bibitem[\protect\citeauthoryear{{Libeskind}, {Hoffman}, {Forero-Romero},
  {Gottl{\"o}ber}, {Knebe}, {Steinmetz} \& {Klypin}}{{Libeskind}
  et~al.}{2013}]{libeskind13a}
{Libeskind} N.~I.,  {Hoffman} Y.,  {Forero-Romero} J.,  {Gottl{\"o}ber} S.,
  {Knebe} A.,  {Steinmetz} M.,    {Klypin} A.,  2013, \mnras, 428, 2489

\bibitem[\protect\citeauthoryear{{Libeskind}, {Hoffman}, {Knebe}, {Steinmetz},
  {Gottl{\"o}ber}, {Metuki} \& {Yepes}}{{Libeskind} et~al.}{2012}]{libeskind12}
{Libeskind} N.~I.,  {Hoffman} Y.,  {Knebe} A.,  {Steinmetz} M.,
  {Gottl{\"o}ber} S.,  {Metuki} O.,    {Yepes} G.,  2012, \mnras, 421, L137

\bibitem[\protect\citeauthoryear{{Libeskind}, {Hoffman}, {Steinmetz},
  {Gottl{\"o}ber}, {Knebe} \& {Hess}}{{Libeskind} et~al.}{2013}]{libeskind13b}
{Libeskind} N.~I.,  {Hoffman} Y.,  {Steinmetz} M.,  {Gottl{\"o}ber} S.,
  {Knebe} A.,    {Hess} S.,  2013, \apjl, 766, L15

\bibitem[\protect\citeauthoryear{{Ocvirk}, {Pichon} \& {Teyssier}}{{Ocvirk}
  et~al.}{2008}]{ocvirk08}
{Ocvirk} P.,  {Pichon} C.,    {Teyssier} R.,  2008, \mnras, 390, 1326

\bibitem[\protect\citeauthoryear{Paz, Stasyszyn \& Padilla}{Paz
  et~al.}{2008}]{Paz08}
Paz D.~J.,  Stasyszyn F.,    Padilla N.~D.,  2008, \mnras, 389, 1127P

\bibitem[\protect\citeauthoryear{{Peebles}}{{Peebles}}{1969}]{peebles69}
{Peebles} P.~J.~E.,  1969, \apj, 155, 393

\bibitem[\protect\citeauthoryear{{Peirani}, {Mohayaee} \& {de Freitas
  Pacheco}}{{Peirani} et~al.}{2004}]{Peirani2004}
{Peirani} S.,  {Mohayaee} R.,    {de Freitas Pacheco} J.~A.,  2004, \mnras,
  348, 921

\bibitem[\protect\citeauthoryear{{Pichon} \& {Bernardeau}}{{Pichon} \&
  {Bernardeau}}{1999}]{pichon99}
{Pichon} C.,  {Bernardeau} F.,  1999, \aap, 343, 663

\bibitem[\protect\citeauthoryear{{Pichon}, {Codis}, {Pogosyan}, {Dubois},
  {Desjacques} \& {Devriendt}}{{Pichon} et~al.}{2014}]{pichon2014}
{Pichon} C.,  {Codis} S.,  {Pogosyan} D.,  {Dubois} Y.,  {Desjacques} V.,
  {Devriendt} J.,  2014, ArXiv e-prints

\bibitem[\protect\citeauthoryear{{Pichon}, {Pogosyan}, {Kimm}, {Slyz},
  {Devriendt} \& {Dubois}}{{Pichon} et~al.}{2011}]{pichon11}
{Pichon} C.,  {Pogosyan} D.,  {Kimm} T.,  {Slyz} A.,  {Devriendt} J.,
  {Dubois} Y.,  2011, \mnras, pp 418, 2493

\bibitem[\protect\citeauthoryear{{Press} \& {Schechter}}{{Press} \&
  {Schechter}}{1974}]{PS1974}
{Press} W.~H.,  {Schechter} P.,  1974, \apj, 187, 425

\bibitem[\protect\citeauthoryear{{Prieto}, {Jimenez} \& {Haiman}}{{Prieto}
  et~al.}{2013}]{Prieto2013}
{Prieto} J.,  {Jimenez} R.,    {Haiman} Z.,  2013, \mnras, 436, 2301

\bibitem[\protect\citeauthoryear{{Pueblas} \& {Scoccimarro}}{{Pueblas} \&
  {Scoccimarro}}{2009}]{Pueblas:2008de}
{Pueblas} S.,  {Scoccimarro} R.,  2009, \prd, 80, 043504

\bibitem[\protect\citeauthoryear{{Schaefer}}{{Schaefer}}{2009}]{schaefer08}
{Schaefer} B.~M.,  2009, International Journal of Modern Physics D, 18, 173

\bibitem[\protect\citeauthoryear{{Sousbie}}{{Sousbie}}{2011}]{sousbie111}
{Sousbie} T.,  2011, \mnras, 414, 350

\bibitem[\protect\citeauthoryear{{Sousbie}, {Colombi} \& {Pichon}}{{Sousbie}
  et~al.}{2009}]{sousbie09}
{Sousbie} T.,  {Colombi} S.,    {Pichon} C.,  2009, \mnras, 393, 457

\bibitem[\protect\citeauthoryear{{Sousbie}, {Pichon} \& {Kawahara}}{{Sousbie}
  et~al.}{2011}]{sousbie112}
{Sousbie} T.,  {Pichon} C.,    {Kawahara} H.,  2011, \mnras, 414, 384

\bibitem[\protect\citeauthoryear{{Spergel}, {Verde}, {Peiris}, {Komatsu},
  {Nolta}, {Bennett}, {Halpern}, {Hinshaw}, {Jarosik}, {Kogut}, {Limon},
  {Meyer}, {Page}, {Tucker}, {Weiland}, {Wollack} \& {Wright}}{{Spergel}
  et~al.}{2003}]{Spergeletal03}
{Spergel} D.~N.,  {Verde} L.,  {Peiris} H.~V.,  {Komatsu} E.,  {Nolta} M.~R.,
  {Bennett} C.~L.,  {Halpern} M.,  {Hinshaw} G.,  {Jarosik} N.,  {Kogut} A.,
  {Limon} M.,  {Meyer} S.~S.,  {Page} L.,  {Tucker} G.~S.,  {Weiland} J.~L.,
  {Wollack} E.,    {Wright} E.~L.,  2003, \apjs, 148, 175

\bibitem[\protect\citeauthoryear{{Springel}, {Yoshida} \& {White}}{{Springel}
  et~al.}{2001}]{GADGET}
{Springel} V.,  {Yoshida} N.,    {White} S.~D.~M.,  2001, New Astronomy, 6, 79

\bibitem[\protect\citeauthoryear{{Stewart}, {Brooks}, {Bullock}, {Maller},
  {Diemand}, {Wadsley} \& {Moustakas}}{{Stewart} et~al.}{2013}]{Stewart2013}
{Stewart} K.~R.,  {Brooks} A.~M.,  {Bullock} J.~S.,  {Maller} A.~H.,  {Diemand}
  J.,  {Wadsley} J.,    {Moustakas} L.~A.,  2013, \apj, 769, 74

\bibitem[\protect\citeauthoryear{{Sutherland} \& {Dopita}}{{Sutherland} \&
  {Dopita}}{1993}]{sutherland&dopita93}
{Sutherland} R.~S.,  {Dopita} M.~A.,  1993, \apjs, 88, 253

\bibitem[\protect\citeauthoryear{Tempel, Stoica \& Saar}{Tempel
  et~al.}{2013}]{Tempel13}
Tempel E.,  Stoica R.~S.,    Saar E.,  2013, Monthly Notices of the Royal
  Astronomical Society, 428, 1827

\bibitem[\protect\citeauthoryear{{Teyssier}}{{Teyssier}}{2002}]{teyssier02}
{Teyssier} R.,  2002, \aap, 385, 337

\bibitem[\protect\citeauthoryear{{Teyssier}, {Pires}, {Prunet}, {Aubert},
  {Pichon}, {Amara}, {Benabed}, {Colombi}, {Refregier} \& {Starck}}{{Teyssier}
  et~al.}{2009}]{Teyssier2009}
{Teyssier} R.,  {Pires} S.,  {Prunet} S.,  {Aubert} D.,  {Pichon} C.,  {Amara}
  A.,  {Benabed} K.,  {Colombi} S.,  {Refregier} A.,    {Starck} J.-L.,  2009,
  \aap, 497, 335

\bibitem[\protect\citeauthoryear{{Tillson}, {Devriendt}, {Slyz}, {Miller} \&
  {Pichon}}{{Tillson} et~al.}{2012}]{tillson12}
{Tillson} H.,  {Devriendt} J.,  {Slyz} A.,  {Miller} L.,    {Pichon} C.,  2012,
  ArXiv:1211.3124

\bibitem[\protect\citeauthoryear{{Wang}, {Szalay}, {Arag{\'o}n-Calvo},
  {Neyrinck} \& {Eyink}}{{Wang} et~al.}{2014}]{Wang13}
{Wang} X.,  {Szalay} A.,  {Arag{\'o}n-Calvo} M.~A.,  {Neyrinck} M.~C.,
  {Eyink} G.~L.,  2014, \apj, 793, 58

\bibitem[\protect\citeauthoryear{White}{White}{1984}]{TTT}
White S. D.~M.,  1984, \apj, 286, 38

\bibitem[\protect\citeauthoryear{{Zhang}, {Yang}, {Faltenbacher}, {Springel},
  {Lin} \& {Wang}}{{Zhang} et~al.}{2009}]{Zhang2009}
{Zhang} Y.,  {Yang} X.,  {Faltenbacher} A.,  {Springel} V.,  {Lin} W.,
  {Wang} H.,  2009, \apj, 706, 747

\bibitem[\protect\citeauthoryear{{Zhang}, {Yang}, {Wang}, {Wang}, {Mo} \& {van
  den Bosch}}{{Zhang} et~al.}{2013}]{Zhang2013}
{Zhang} Y.,  {Yang} X.,  {Wang} H.,  {Wang} L.,  {Mo} H.~J.,    {van den Bosch}
  F.~C.,  2013, \apj, 779, 160

\end{thebibliography}

%%%%%%%%%%%%%%%%%%%
%%%%%%%%%%%%%%%%%%%

\appendix

\section{ The vorticity of the gas }
 \label{sec:gas-LSS}
\label{sec:gas}

%%%%%%%%%
\begin{figure}
\begin{center}
\includegraphics[scale=0.55]{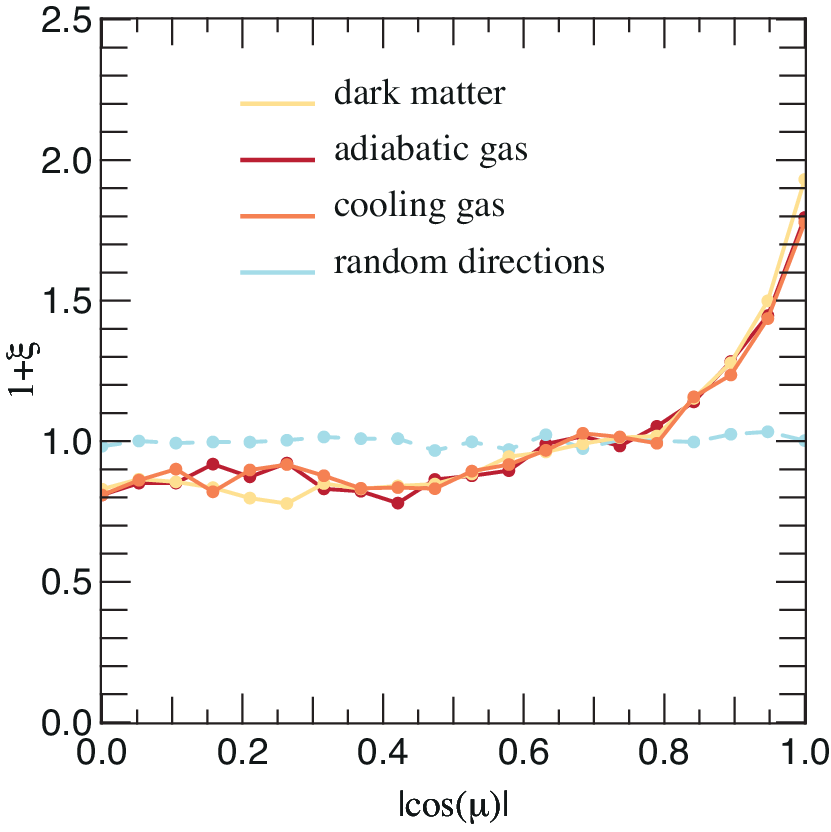} 
\end{center}
\caption{
The probability distribution of the cosine of the angle between the vorticity in the smoothed DM 
 and hydrodynamical simulations  and the direction of the filament (solid line). The same measure is done for random directions of $u$ (dashed blue line), plotted for the DM in ${\cal S}^{\rm HDM}_{100}$ and for the gas in ${\cal S}^{\rm HA}_{100}$ (adiabatic gas) and ${\cal S}^{\rm HC}_{100}$ (cooling run). We find an excess of probability for $\vert\cos \mu\vert$ in $\left[ 0.5,1\right]$ relative to random orientations, and three profiles are very similar, which shows that large-scale modes dominate.
 }
\label{fig:cosinePDF-hydro}
\end{figure}
%%%%%%%%%

\begin{table}
\singlespacing
\begin{tabular}{l|l|l|c}
 \textbf{Name} & \textbf{Type} & \textbf{Box size}& \textbf{Resolution}\\
  & & $h^{-1}\, \rm Mpc$ & \\\hline
 ${\cal S}^{\rm HA}_{100}$ & $\Lambda$HDM\, adiabatic &100 & $256^3$ \\
${\cal S}^{\rm HC}_{100}$ & $\Lambda$HDM\, cool &100 & $256^3$ \\
${\cal S}^{\rm cool}_{20}(0.7)$ & $\Lambda$CDM\, cool& \,\,20 & $1024^3$ \\
\end{tabular}
\caption{The set of hydrodynamical simulations used in Appendix~\ref{sec:gas}.
 The hydro runs come in two categories: adiabatic and cooling, including one high resolution run
which was stopped at redshift 0.7.  
}
\label{tab:simuhydro}
\end{table}

We use  three hydrodynamical simulations
 ${\cal S}^{\rm HA}_{100}$,
${\cal S}^{\rm HC}_{100}$ and
${\cal S}^{\rm cool}_{20}(0.7)$ (see also Table~\ref{tab:simuhydro}),  carried out with the Eulerian hydrodynamic code {\sc ramses}~\citep{teyssier02}, which uses an Adaptative Mesh Refinement (AMR) technique. 
For these hydrodynamical runs, the evolution of the gas is followed using a second-order unsplit Godunov scheme for the Euler equations. 
The HLLC Riemann solver with a first-order MinMod Total Variation Diminishing scheme to reconstruct the interpolated variables from their cell-centred values is used to compute fluxes at cell interfaces. 
Collisionless particles (DM and star particles) are evolved using a particle-mesh solver with a Cloud-In-Cell interpolation.
The initial mesh is refined up to $\Delta x=1.7\, \rm kpc$  according to a quasi-Lagragian criterion: if the number of DM particles in a cell is more than eight, or if the total baryonic mass in a cell is eight times the initial DM mass resolution.

For the cooling runs ${\cal S}^{\rm cool}_{100}$, and ${\cal S}^{\rm cool}_{20}$, gas is allowed to cool by H and He cooling with an eventual contribution from metals using a~\cite{sutherland&dopita93} model down to $10^4\, \rm K$.
Heating from a uniform UV background takes place after redshift $z_{\rm reion} = 10$ following~\cite{haardt&madau96}.

On large scales (as probed by the smoothed sets of simulations) the vorticity of  gas shows the same correlations with the filaments as DM does. 
Fig.~\ref{fig:cosinePDF-hydro} displays the probability distribution of the cosine of the angle between the vorticity  and the direction of the filament for the DM field (in red), the adiabatic gas (in blue) cooling run (in yellow).
These three simulations quantitatively show the same preference for their vorticity to be aligned with the filamentary structure.
In a nutshell, differences between the adiabatic and the cooling run only appear on kpc scales, so that on large scales, the DM, adiabatic and cooling runs have the same velocity field structure.

Fig.~\ref{fig:cool-redshift} displays the  probability distribution of the cosine of the angle between the vorticity and the direction of the skeleton
for a range of redshifts.  The correlation between the direction of the filament and the vorticity is significant.
As expected, this correlation decreases with cosmic time (at a fixed smoothing scale).
Appendix~\ref{sec:persistence} investigates the evolution of this correlation as a function of the skeleton's persistence. 
As long as we consider large enough scales, the alignment pervades and is consistent with that of the DM.  
 On smaller scales, the gas is dense enough to allow cooling to operate and
re-structure the velocity flow. Notwithstanding, these smaller scale structures do not affect the larger scale correlation between vorticity and the direction of the filaments. 

%%%%%%%%%%%%%%%%%%%%%%%

\section{Toy model for halo spin} \label{sec:Toy}

Can a model based on a vorticity field in qualitative agreement with what was found in the simulation explain why it should lead the observed evolution of 
spin alignment with mass?

\begin{figure*}
\begin{center}
\includegraphics[scale=0.6]{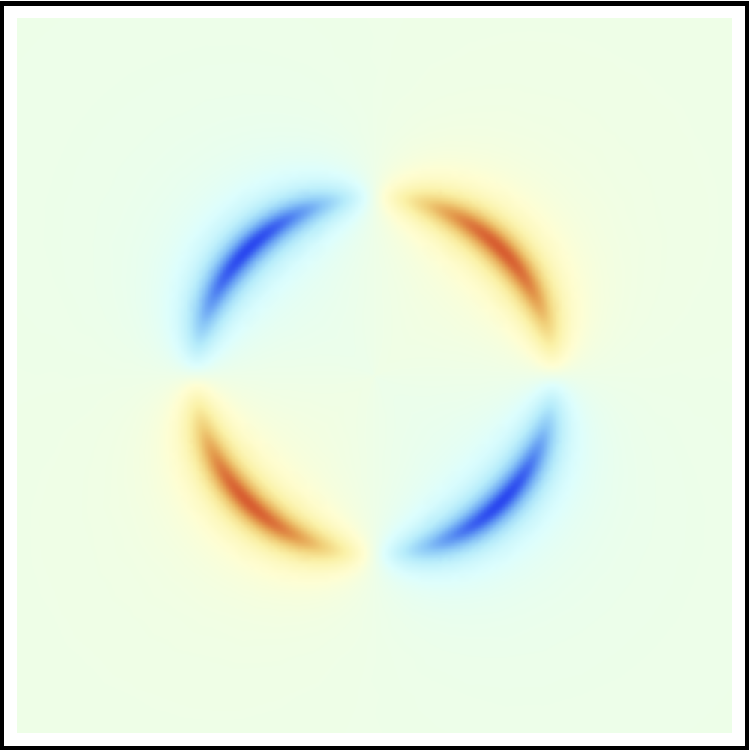}
\includegraphics[scale=0.6]{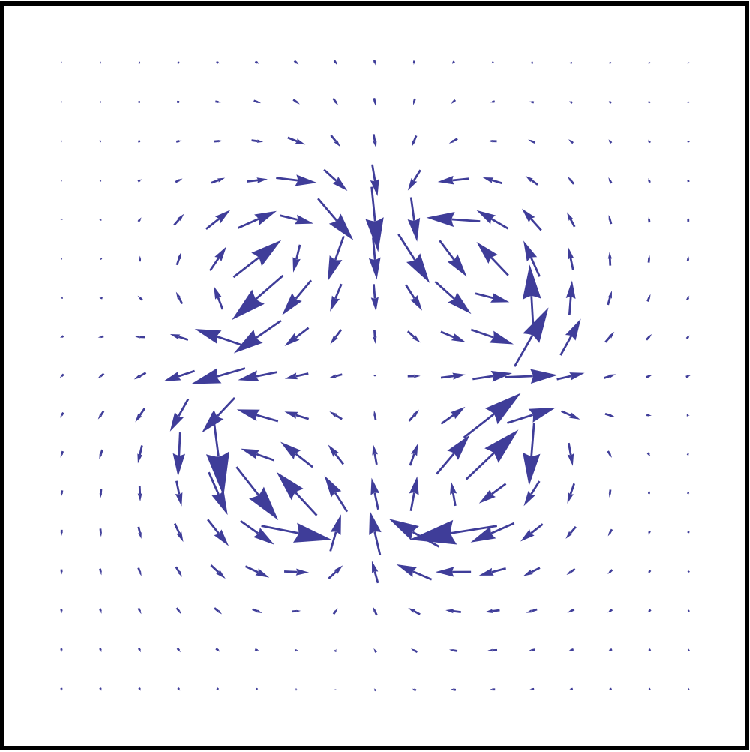}
\includegraphics[scale=0.6]{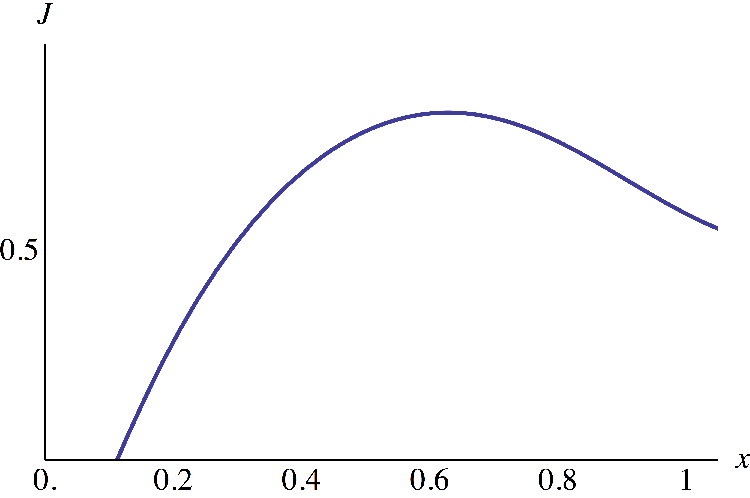}
\includegraphics[scale=0.6]{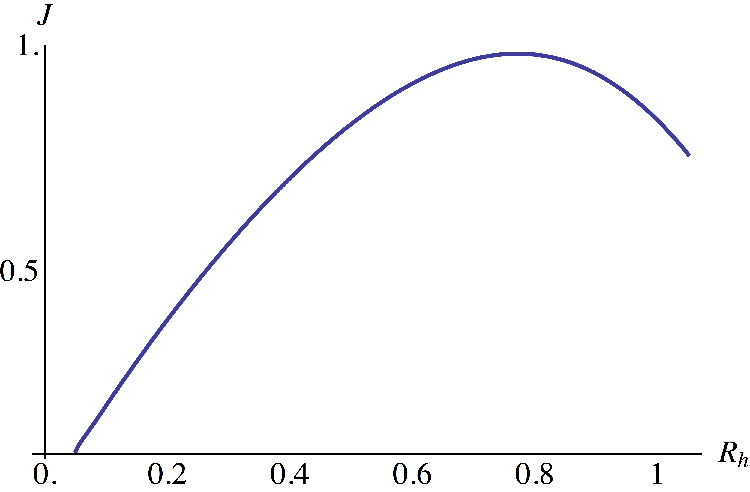}
\end{center}

\caption{  {\sl Top:} maps of the vorticity ({\sl left})  projected along $\mathbf{e_{z}}$ and the associated curl component of the velocity ({\sl right}) computed from the vorticity by application of the Helmoltz-Hodge theorem. {\sl Bottom:} magnitude of the AM along $\mathbf{e_{z}}$ for a halo embedded in one of the four quadrants. We first consider how it varies as a function of the radius of the halo ({\sl left}), the position of this latter being fixed ($x_{\rm h}$,$y_{\rm h}$)=(0.5/$\sqrt{2}$ , 0.5/$\sqrt{2}$). The alignment of the AM of the halo with the vorticity increases
until the halo size becomes comparable to that of the vorticity quadrant. We study then the magnitude of the AM along $\mathbf{e_{z}}$ at fixed radius ($R_{h}=0.3$) as a function of the position $x$ along the diagonal ({\sl right}). In this case, the alignment increases up to the point where the halo boundary coincides
with the vorticity caustic, with the halo still being fully contained within the vorticity quadrant ($x_{\rm h}$,$y_{\rm h}$)$\sim$(0.7/$\sqrt{2}$ , 0.7/$\sqrt{2}$).} \label{fig:spin}
\end{figure*}

Let us qualitatively illustrate with a simple toy model  this mass transition for the spin-filament alignment.
 In this toy problem, we consider an isolated infinite filament aligned along $\mathbf{e_{\rm z}}$. We define the corresponding
 idealized  vorticity field as  \[
 {\mathbf \Omega}(\mathbf{r},\theta) ={\rm C}{\epsilon}\sin(2\theta)\frac{1}{(\epsilon^{2}+(r-R)^{2})}\,\mathbf{e_{\rm z}}\,,
 \]  with ${\rm C}$ a constant, $R$ the radius of the caustic and $\epsilon$ a small number. The vorticity thus defined is largest along the caustic,
 point reflection symmetric and tends rapidly to $0$ outside the caustic. Should $\epsilon$ tend to $0$, vorticity would become singular on the edge of the caustic ($r \rightarrow R$). The map of the vorticity is displayed in Fig.~\ref{fig:spin} (top left panel).

By application of the Helmoltz-Hodge theorem, we find that the curl component of the velocity field consistent  with that vorticity (i.e. such that $
{\mathbf \Omega} =\nabla\times\mathbf{v}$) obeys
\begin{equation}
\mathbf{v}(\mathbf{r},\theta)=\dfrac{1}{4\pi}\int_{{\cal V}}{\nabla\times\mathbf{\Omega}(\mathbf{r'},\theta')}\frac{1}{\vert \mathbf{r}-\mathbf{r'}\vert}{\rm d}\cal V \,. \label{eq:vel-vort}
\end{equation} 
We assume here that the shear part of the curl free component of the velocity flow is smaller on scales comparable to the halo.
We now consider  a spherical halo of radius $r_{{\rm h}}$ embedded in one of the four quadrants of the caustic, centred on $C_{{\rm h}} (x_{\rm h},y_{\rm h}$) with $x_{\rm h}^{2}+y_{\rm h}^{2}\le R$. From equation (~\ref{eq:vel-vort}) we can simply compute its AM ${\cal J}(r_{\rm h},x_{\rm h},y_{\rm h})$, and look at the variation of ${\cal J}$ as a function of its position at fixed radius, or as a function of its radius at fixed position. Fig.~\ref{fig:spin} (bottom panel) shows the magnitude of the AM along the $z$-axis for a halo centred on $C_{{\rm h}}$(0.5 R/$\sqrt{2}$,0.5 R/$\sqrt{2}$) as a function of the radius. We observe that the alignment increases until the size of the halo encompasses the whole quadrant. At a given radius, the position of the halo which maximizes the AM is the one for which the edge of the halo coincides with the edge of the caustic, since the vorticity peaks close to the caustic.

%%%%%%%%%%%%%%%%%%%%%%%

\section{Persistence effects} \label{sec:validation-persistence}
%%%%%%%%%%%%%%%%%%%
\label{sec:persistence}

Given the characteristics  of $\Lambda$CDM hierarchical clustering,
one can anticipate that the process described in the main text occurs on several nested scales at various epochs -- and arguably  on various scales at the same epoch. The scenario we propose for the origin of vorticity and  spin alignment is, like the signal itself, relative to the linear scale involved in defining the filaments and as such, multi-scale.
Indeed in the main text, the two sets of simulations, ${\cal S}^{\Lambda{\rm CDM}}$ and ${\cal S}^{\Lambda{\rm HDM}}$, allowed us to 
probe different scales of the vorticity field. 
The induced multi-scale anisotropic flow also transpires in the scaling of the spin flipping transition mass with smoothing presented in apppendix~D of \cite{codis12}.
 It will hold as long as filaments are well defined in order to drive the local cosmic flow.  

%%%%%%%%%
\begin{figure*}\begin{center}
\includegraphics[scale=0.55]{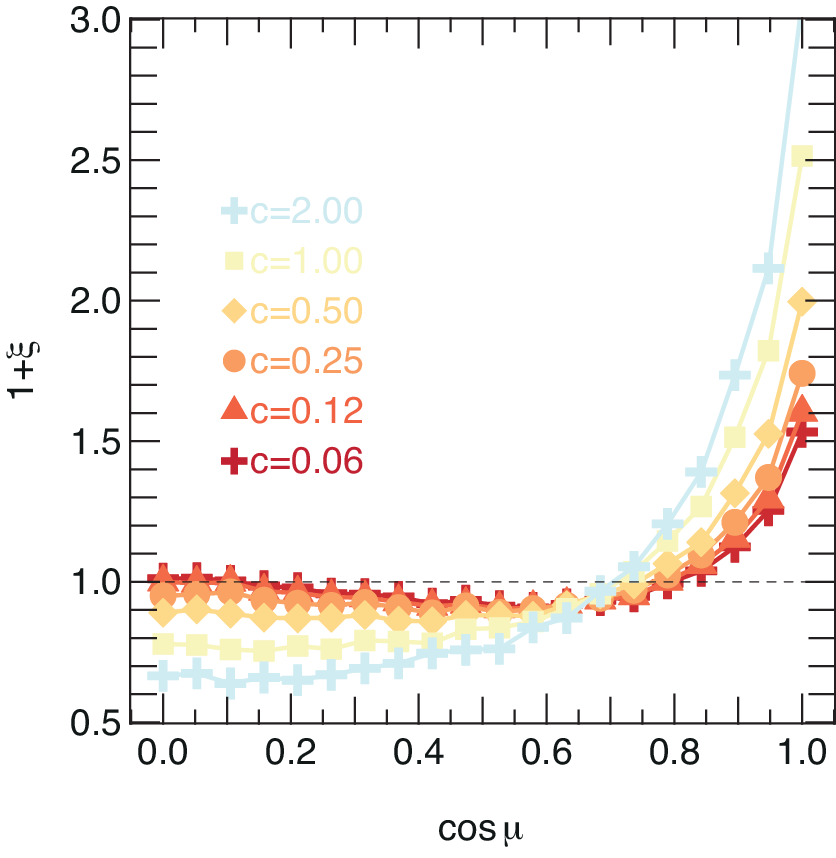} 
\includegraphics[scale=0.575]{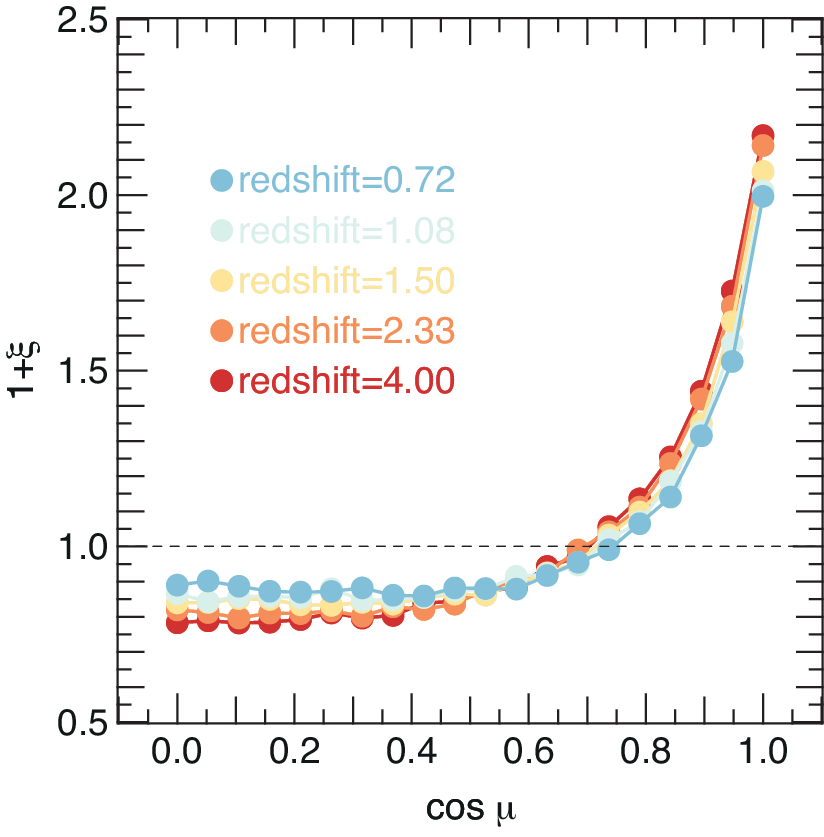}
\end{center}
\caption{{\sl Left} The probability distribution of the cosine of the angle between the vorticity and the direction of the filament measured in ${\cal S}^{\rm cool}_{20}(z=0.7)$
for different persistence threshold. The level of persistence of the main text corresponds to $c=0.5$.
{\sl Right}
The probability distribution of the cosine of the angle between the vorticity and the direction of the skeleton, measured in ${\cal S}^{\rm cool}_{20}(0.7)$ for various redshifts as labeled. The  amplitude of the correlation decreases with cosmic time.}
\label{fig:cool-redshift}\label{fig:align-cool}
\end{figure*}
%%%%%%%%%

Let us now briefly explore the effects of probing different scales of the LSS via the skeleton level of persistence. 
Fig.~\ref{fig:align-cool} shows the excess alignment probability as a function of the cosine of the angle between the vorticity and the filaments 
as a function of the persistence level for a range of values. The alignment is strongest with the largest scale filamentary structure corresponding to the least dynamically evolved features of the field.
Here the gas density was sampled over a cube of size $512^3$. It was then 
smoothed over 8 pixels (300 kpc) and the persistent skeleton was computed from the logarithm of that smoothed field normalized to its standard 
deviation.  Hence the persistence levels 0.06, 0.12, ..., 2 are in units of this root mean square.

%%%%%%%%%
\begin{figure}\begin{center}
\includegraphics[scale=0.425]{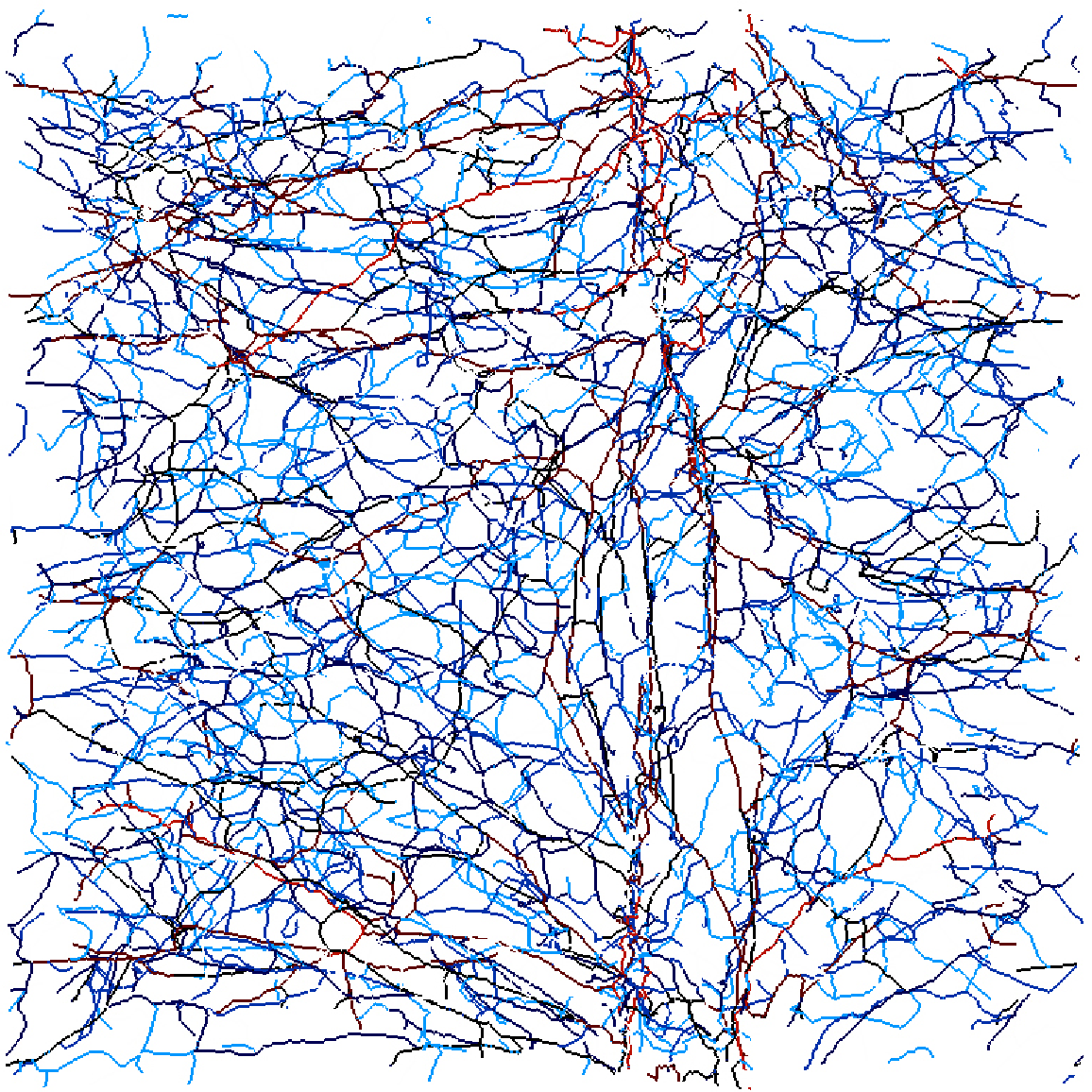} 
\end{center}
\caption{The skeleton measured in ${\cal S}^{\rm cool}_{20}(0.7)$
for increasing persistence threshold, 0.06, 0.12, ..., 2, from light blue to red; the 
skeleton has tree-like structure where the main branches correspond to the most persistent ones.
The level of persistence of the main text corresponds to the dark blue and red branches.}\label{fig:skel-cool}
\end{figure}
%%%%%%%%%

Fig.~\ref{fig:skel-cool} gives visual impression of the corresponding structure of the skeleton as a function of these persistence levels:
the skeleton has a tree-like structure, for which each level of lower persistence contributes smaller branches. 
Hence the persistence level of 0.5 used in the main text corresponds to a description of the main filaments of the simulation.

\section{ The effect of  smoothing} \label{sec:validation-smoothing}
%%%%%%%%%%%%%%%%%%%
\label{sec:smoothing}

%%%%%%%%%
\begin{figure}\begin{center}
\includegraphics[scale=0.55]{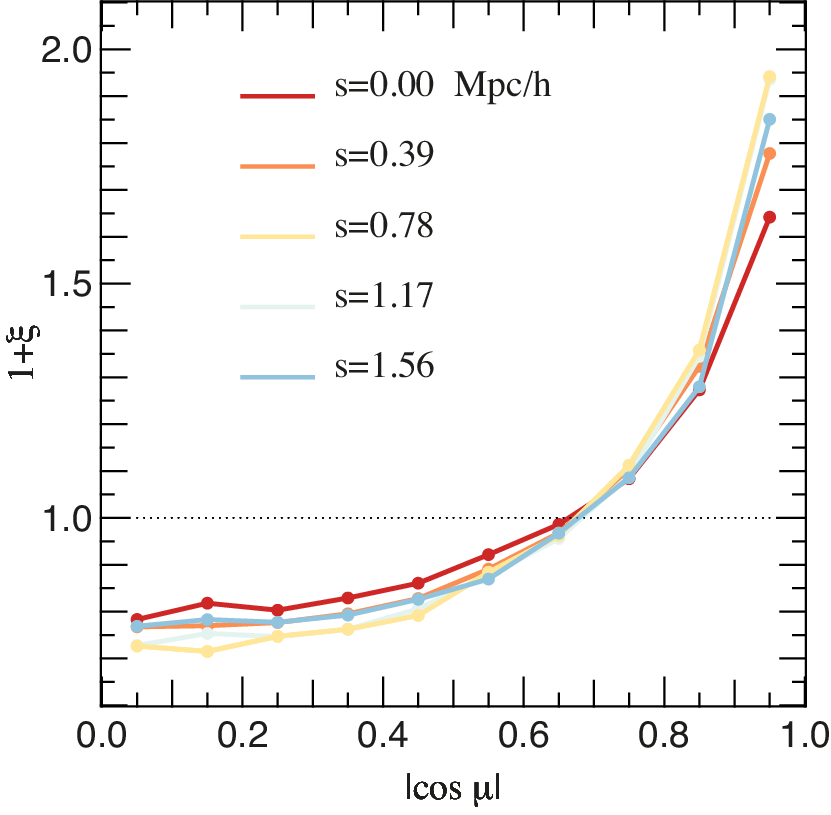} 
\end{center}
\caption{The probability distribution of the cosine of the angle between the vorticity and the direction of the filament, measured in ${\cal S}^{\rm CDM}_{100}$ for various smoothing scales of the velocity field before computation of the vorticity. Smoothing scales are expressed in $h^{-1}$ Mpc. The smoothing scale adopted in the main text is 0.39 $h^{-1}$ Mpc.} \label{fig:align-var-smooth}
\end{figure}
%%%%%%%%%

Fig.~\ref{fig:align-var-smooth} shows the effect of the smoothing of the velocity field before computing the vorticity, on the alignment between the vorticity and the direction of the filament. The amplitude of the excess of alignment varies slightly with the smoothing scale, but the main conclusion that an excess of alignment is detected remains unchanged.

%%%%%%%%%%%%%%%%%%%%%%%

\section{Tidal -vorticity lockup} \label{sec:validation}
\begin{figure}\begin{center}
\includegraphics[scale=0.55]{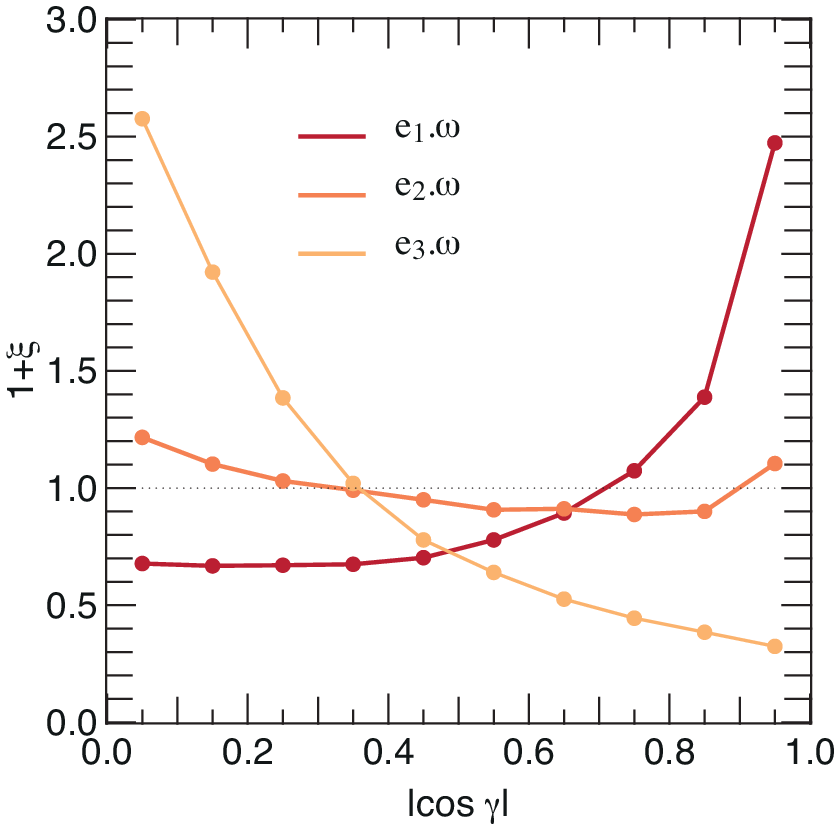} 
\end{center}
\caption{The probability distribution of the cosine of the angle between the vorticity and the eigenvectors of the tidal tensor, measured in ${\cal S}^{\rm CDM}_{100}$. The vorticity tends to be perpendicular to the minor axis ($e_{3}$) of the tidal tensor: 
the excess of probability to have $\vert\cos\theta \vert$ in $\left[ 0, 0.5\right]$ (i.e. $60\leq\theta\leq 90^{\rm o}$) is 50 per cent relative to random orientations. The vorticity tends also to be aligned with the major axis ($e_{1}$): the excess probability to have $\vert\cos\theta \vert$ in $\left[ 0.5, 1\right]$ (i.e. $0\leq\theta\leq 60^{\rm o}$) is 25 per cent relative to random orientations. $e_{3}$ corresponds to the axis along which material is collapsing fastest.
}\label{fig:sheartensor}
\end{figure}

%%%%%%%%%%%%%%%%%%%
%
Fig.~\ref{fig:sheartensor} displays  the probability distribution of the cosine of the angle between the vorticity and the eigenvectors of the tidal field tensor, $\cos \gamma$.
The vorticity tends to be perpendicular to the minor axis ($e_{3}$) of the tidal tensor which corresponds to the axis along which material is collapsing fastest. It is qualitatively in agreement with Fig.~\ref{fig:cosinePDF} and with \cite{libeskind13b} which focus, respectively, on the eigenvectors of the Hessian
of the density, and the eigenvectors of the shear tensor. For the latter, the description is kinematic, rather than dynamical  for the tidal field.

\section{Fof halo catalogue } \label{sec:Fof}
%%%%%%%%%%%%%%%%%%%
%
As mentioned in the main text,  FOF is prone to spuriously link neighbouring structures which could bias the alignment of the spin and the vorticity. 
An additional criterion is therefore required to produce a trustworthy catalogue of haloes. Following \cite{Bett21032007}, we proceed using  the distribution of the spin parameter defined by \cite{peebles69}: $\lambda={J\vert E \vert^{1/2}}/{{\rm G} M_{h}^{5/2}}$, where $J$ is the magnitude of the spin, $E$ is the total energy of the halo, G is the gravitational constant and $M_{h}$ is the halo mass. 
 
\begin{figure}\begin{center}
\includegraphics[scale=0.55]{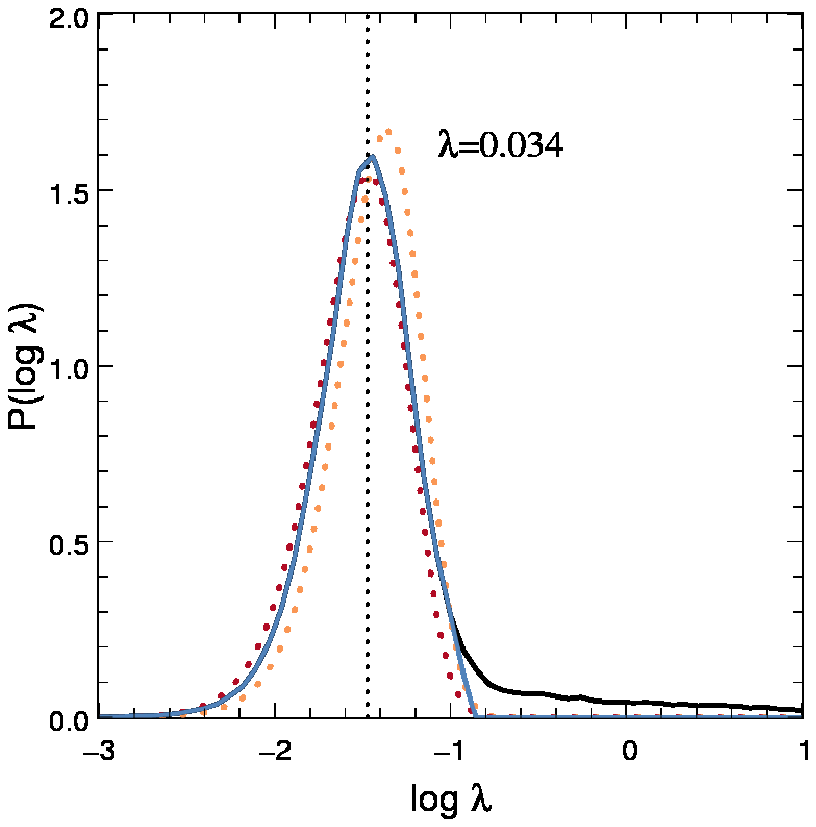} 
\end{center}
\caption{The normalized histogram of the logarithm of the spin parameter in the simulations set ${\cal S}^{\rm CDM}_{50}$. The median value of $\lambda$ is 0.034 $\pm$ 0.0005 for the cleaned catalog. The solid black line is the normalized distribution for all haloes. Notice a tail at high spin parameter which corresponds to spuriously linked structures. The solid blue line is the distribution for haloes with $\lambda \le 0.12$, with the same normalization as for all haloes. The dotted lines are the analytical fit described in the text, red is our best fit to the distribution, and orange is the best fit found by Bett et al. (2007).}\label{fig:lambdadistrib}
\end{figure}

 Fig.~\ref{fig:lambdadistrib} shows the average normalized histogram of the logarithm of the spin parameter for the haloes in the simulations set ${\cal S}^{\rm CDM}_{50}$. At high spin we clearly see a long tail, up to $\lambda = 238.2$, due to spurious linking of the structures. We use the analytical model proposed in \cite{Bett21032007} to fit the $\log \lambda$-distribution: $P(\log \lambda)={\rm A} ({\lambda}/{\lambda_{0}})^{3}\exp\left[-\alpha(\lambda/\lambda_{0})^{3/\alpha}\right]$, where $A=3 \ln 10\, \alpha^{\alpha-1}/\Gamma(\alpha)$, with the values $\lambda_{0}=0.0341$  and $\alpha=2.98$ which are providing the best fit. These values are in good agreement with those found by \cite{Bett21032007} ($\lambda_{0}=0.043$  and $\alpha=2.51$), though their way to clean their catalogue (TREEall) is more sophisticated, in particular by taking into account an additional  condition on energy. They showed also that the minimal number of particles per halo, $N_{p}$, clearly affects the $\lambda$-distribution only for $N_{p}$ lower than 100. Above this threshold, the change in the median value of $\lambda$ stays lower than 10\%. Consequently, we keep  in our catalogue only haloes with more than 100 particles. These haloes are then selected through a cut in $\lambda$. We find that removing haloes with $\lambda \ge 0.12$ best fits the adopted analytical model. Removed haloes represent $9 .4 \pm 1.2$ \% of the total population. Inspection of some of these removed haloes shows that they generally are multiple objects. We are left with around 5000 haloes in each 50 $h^{-1}\,Ê\rm Mpc$ box of the ${\cal S}^{\rm CDM}_{50}$ simulations set. 

We then quantify how the cut in $\lambda$ affects the vorticity--spin alignment results. Considering three different catalogues with three different cuts in $\lambda$ ($\lambda<0.08$, $\lambda<0.12$, $\lambda<0.2$) we look for each catalogue and for each bin of mass at the quantity $(\zeta_{\rm tot}-\zeta_{\rm cut})/(1+\zeta_{\rm tot})$ where $1+\zeta_{\rm cut}$ is the excess of alignment in the reduced  catalogue  and $1+\zeta_{\rm tot}$ in the full catalogue. This difference is always $< 5$\%. We conclude that including or not the misidentified structures does not significantly change the measure of the spin alignment with the vorticity.

%%%%%%%%%%%%%%%%%%%%%%%

%
\section{Zoology of caustics } 
%%%%%%%%%%%%%%%%%%%
%
%

\begin{figure}
\includegraphics[scale=0.3]{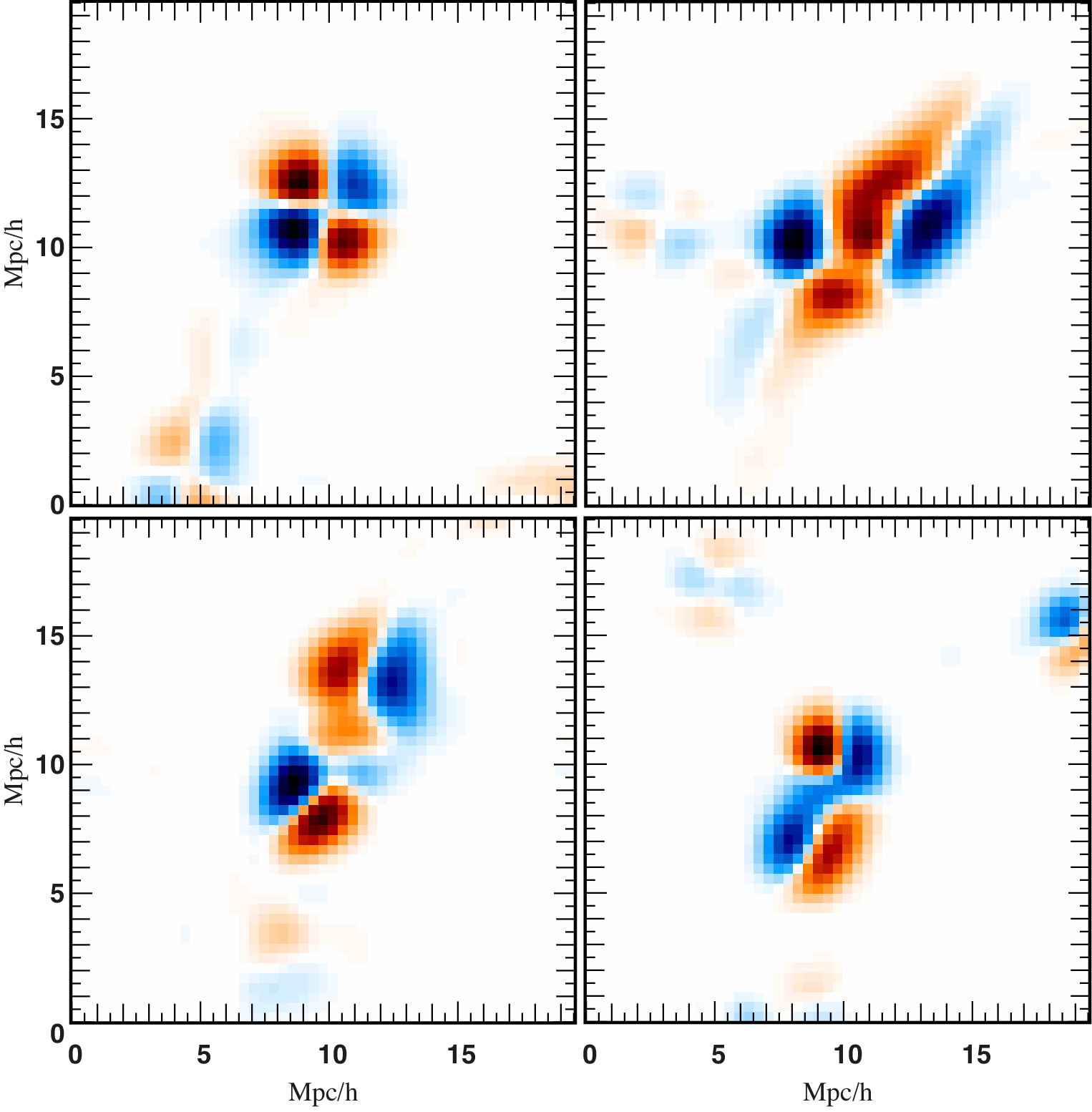}
\caption{Different kinds of vorticity cross-sections.} \label{fig:multisections}
\end{figure}

Fig.~\ref{fig:multisections} shows a bundle of cross-sections of vorticity computed as in Fig.~\ref{fig:vorticity}.
%%%%%%%%%%%%%%%%%%%%%%%
\section{Defining zero vorticity }
\label{sec:walls}

The algorithm {\sc DisPerSE} introduced by \cite{sousbie111} is used to defined the density walls and the contours of minimal vorticity. The density walls are computed as being the ascending two--manifolds of the skeleton calculated on the density field. The contours of minimal vorticity are defined as being the descending two--manifolds of the skeleton calculated on the norm of the vorticity field. Since the vorticity is really well defined only on the neighbourhood of caustics, a mask is applied when the walls are computed, which covers all the regions of space where the density is lower than 10 per cent of the maximum density and the vorticity lower than 10 per cent of the maximum vorticity. The results of the computation of the density walls and minimal vorticity contours are tessellations, which means sets of triangles. For each triangle in the minimal vorticity tessellation we find its nearest 
neighbours in the density tessellation. Smoothing is achieved by averaging the position of each vertex with that of its direct neighbours. A smoothing coefficient $S=$N means that this operation is repeated $N$ times. The cosine between the normals of both triangles is then calculated.

\end{document}